\documentclass[prc,twocolumn,superscriptaddress,showpacs,twoside,floatfix]
{revtex4-1}

\usepackage{amssymb,epsfig}
\usepackage{color}
\begin{document}

\title{Deep excursion beyond the proton dripline. II. Toward the limits of existence of nuclear structure}

\author{L.V.~Grigorenko}
\affiliation{Flerov Laboratory of Nuclear Reactions, JINR,  141980 Dubna, Russia}
\affiliation{National Research Nuclear University ``MEPhI'',
115409 Moscow, Russia}
\affiliation{National Research Centre ``Kurchatov Institute'', Kurchatov
sq.\ 1, 123182 Moscow, Russia}

\author{I.~Mukha}
\affiliation{GSI Helmholtzzentrum  f\"{u}r Schwerionenforschung GmbH, 64291
Darmstadt, Germany}

\author{D.~Kostyleva}
\email{D.Kostyleva@gsi.de}
\affiliation{II.Physikalisches Institut, Justus-Liebig-Universit\"at, 35392 Gie{\ss}en, Germany}
\affiliation{GSI Helmholtzzentrum  f\"{u}r Schwerionenforschung GmbH, 64291
Darmstadt, Germany}

\author{C.~Scheidenberger}
\affiliation{GSI Helmholtzzentrum  f\"{u}r Schwerionenforschung GmbH, 64291 Darmstadt, Germany}
\affiliation{II.Physikalisches Institut, Justus-Liebig-Universit\"at, 35392 Gie{\ss}en, Germany}

\author{L.~Acosta}
\affiliation{INFN, Laboratori Nazionali del Sud,
Via S.~Sof\'ia, 95123 Catania, Italy}
\affiliation{Instituto de F\'isica, Universidad Nacional Aut\'onoma de M\'exico, M\'exico, D.F.\ 01000, Mexico}

\author{E.~Casarejos}
\affiliation{University of  Vigo, 36310 Vigo, Spain}

\author{V.~Chudoba}
\affiliation{Flerov Laboratory of Nuclear Reactions, JINR,  141980 Dubna, Russia}
\affiliation{Institute of Physics, Silesian University Opava, 74601 Opava, Czech Republic}

\author{A.A.~Ciemny}
\affiliation{Faculty of Physics, University of Warsaw, 02-093 Warszawa, Poland}

\author{W.~Dominik}
\affiliation{Faculty of Physics, University of Warsaw, 02-093 Warszawa, Poland}

\author{J.A.~Due\~nas}
\affiliation{Depto. de Ingenieria Electrica y Centro de Estudios Avanzados en Fisica, Matem\'{a}ticas y Computaci\'{o}n, Universidad de Huelva, 21071 Huelva, Spain}

\author{V.~Dunin}
\affiliation{Veksler and Baldin Laboratory of High Energy Physics, JINR, 141980 Dubna,
Russia}

\author{J.~M.~Espino}
\affiliation{Department of Atomic, Molecular and Nuclear Physics, University of Seville, 41012 Seville, Spain}

\author{A.~Estrad\'{e}}
\affiliation{University of Edinburgh, EH1 1HT Edinburgh, United Kingdom}

\author{F.~Farinon}
\affiliation{GSI Helmholtzzentrum  f\"{u}r Schwerionenforschung GmbH, 64291 Darmstadt, Germany}

\author{A.~Fomichev}
\affiliation{Flerov Laboratory of Nuclear Reactions, JINR,  141980 Dubna, Russia}

\author{H.~Geissel}
\affiliation{GSI Helmholtzzentrum  f\"{u}r Schwerionenforschung GmbH, 64291 Darmstadt, Germany}
\affiliation{II.Physikalisches Institut, Justus-Liebig-Universit\"at, 35392 Gie{\ss}en, Germany}

\author{A.~Gorshkov}
\affiliation{Flerov Laboratory of Nuclear Reactions, JINR,  141980 Dubna, Russia}

\author{Z.~Janas}
 \affiliation{Faculty of Physics,  University of Warsaw, 02-093 Warszawa,
 Poland}

\author{G.~Kami\'{n}ski}
\affiliation{Heavy Ion Laboratory, University of Warsaw, 02-093 Warszawa,
Poland}
\affiliation{Flerov Laboratory of Nuclear Reactions, JINR,  141980 Dubna,
Russia}

\author{O.~Kiselev}
\affiliation{GSI Helmholtzzentrum  f\"{u}r Schwerionenforschung GmbH, 64291 Darmstadt, Germany}

\author{R.~Kn\"{o}bel}
\affiliation{GSI Helmholtzzentrum  f\"{u}r Schwerionenforschung GmbH, 64291 Darmstadt, Germany}
\affiliation{II.Physikalisches Institut, Justus-Liebig-Universit\"at, 35392 Gie{\ss}en, Germany}

\author{S.~Krupko}
\affiliation{Flerov Laboratory of Nuclear Reactions, JINR,  141980 Dubna, Russia}

\author{M.~Kuich}
\affiliation{Faculty of Physics, Warsaw University of Technology, 00-662 Warszawa, Poland}
 \affiliation{Faculty of Physics,  University of Warsaw, 02-093 Warszawa,
 Poland}

\author{Yu.A.~Litvinov}
\affiliation{GSI Helmholtzzentrum  f\"{u}r Schwerionenforschung GmbH, 64291 Darmstadt, Germany}

\author{G.~Marquinez-Dur\'{a}n}
\affiliation{Department of Applied Physics, University of Huelva, 21071 Huelva, Spain}

\author{I.~Martel}
\affiliation{Department of Applied Physics, University of Huelva, 21071 Huelva, Spain}

\author{C.~Mazzocchi}
\affiliation{Faculty of Physics, University of Warsaw, 02-093 Warszawa, Poland}

\author{E.Yu.~Nikolskii}
\affiliation{National Research Centre ``Kurchatov Institute'', Kurchatov
sq.\ 1, 123182 Moscow, Russia}
\affiliation{Flerov Laboratory of Nuclear Reactions, JINR,  141980 Dubna, Russia}

\author{C.~Nociforo}
\affiliation{GSI Helmholtzzentrum  f\"{u}r Schwerionenforschung GmbH, 64291 Darmstadt, Germany}

\author{A.~K.~Ord\'{u}z}
\affiliation{Department of Applied Physics, University of Huelva, 21071 Huelva, Spain}

\author{M.~Pf\"{u}tzner}
\affiliation{Faculty of Physics, University of Warsaw, 02-093 Warszawa, Poland}
\affiliation{GSI Helmholtzzentrum  f\"{u}r Schwerionenforschung GmbH, 64291 Darmstadt, Germany}

\author{S.~Pietri}
\affiliation{GSI Helmholtzzentrum  f\"{u}r Schwerionenforschung GmbH, 64291 Darmstadt, Germany}

\author{M.~Pomorski}
 \affiliation{Faculty of Physics,  University of Warsaw, 02-093 Warszawa,
 Poland}

\author{A.~Prochazka}
\affiliation{GSI Helmholtzzentrum  f\"{u}r Schwerionenforschung GmbH, 64291 Darmstadt, Germany}

\author{S.~Rymzhanova}
\affiliation{Flerov Laboratory of Nuclear Reactions, JINR,  141980 Dubna, Russia}

\author{A.M.~S\'{a}nchez-Ben\'{i}tez}
\affiliation{Centro de Estudios Avanzados en F\'{i}sica, Matem\'{a}ticas y
Computaci\'{o}n (CEAFMC), Department of Integrated Sciences, University of
Huelva, 21071 Huelva, Spain}

\author{P.~Sharov}
\affiliation{Flerov Laboratory of Nuclear Reactions, JINR,  141980 Dubna, Russia}

\author{H.~Simon}
\affiliation{GSI Helmholtzzentrum  f\"{u}r Schwerionenforschung GmbH, 64291 Darmstadt, Germany}

\author{B.~Sitar}
\affiliation{Faculty of Mathematics and Physics, Comenius University, 84248 Bratislava,
Slovakia}

\author{R.~Slepnev}
\affiliation{Flerov Laboratory of Nuclear Reactions, JINR,  141980 Dubna, Russia}

\author{M.~Stanoiu}
\affiliation{IFIN-HH, Post Office Box MG-6, Bucharest, Romania}

\author{P.~Strmen}
\affiliation{Faculty of Mathematics and Physics, Comenius University, 84248 Bratislava,
Slovakia}

\author{I.~Szarka}
\affiliation{Faculty of Mathematics and Physics, Comenius University, 84248 Bratislava,
Slovakia}

\author{M.~Takechi}
\affiliation{GSI Helmholtzzentrum  f\"{u}r Schwerionenforschung GmbH, 64291 Darmstadt, Germany}

\author{Y.K.~Tanaka}
\affiliation{GSI Helmholtzzentrum  f\"{u}r Schwerionenforschung GmbH, 64291 Darmstadt, Germany}
\affiliation{University of Tokyo, 113-0033 Tokyo, Japan}

\author{H.~Weick}
\affiliation{GSI Helmholtzzentrum  f\"{u}r Schwerionenforschung GmbH, 64291 Darmstadt, Germany}

\author{M.~Winkler}
\affiliation{GSI Helmholtzzentrum  f\"{u}r Schwerionenforschung GmbH, 64291 Darmstadt, Germany}

\author{J.S.~Winfield}
\affiliation{GSI Helmholtzzentrum  f\"{u}r Schwerionenforschung GmbH, 64291 Darmstadt, Germany}

\author{X.~Xu}
\affiliation{School of Physics and Nuclear Energy Engineering, Beihang University, 100191 Beijing, China}
\affiliation{II.Physikalisches Institut, Justus-Liebig-Universit\"at, 35392 Gie{\ss}en, Germany}
\affiliation{GSI Helmholtzzentrum  f\"{u}r Schwerionenforschung GmbH, 64291 Darmstadt, Germany}

\author{M.V.~Zhukov}
\affiliation{Department of Physics, Chalmers University of Technology, S-41296 G\"oteborg, Sweden}

\collaboration{for the Super-FRS Experiment Collaboration}

\date{\today. {\tt File: ar-cl-excur-f2-7-resubmit.tex }}

\begin{abstract}
Prospects of experimental studies of argon and chlorine isotopes located far beyond the proton dripline are studied by using systematics and cluster models. The  deviations from the widespread systematics observed in $^{28,29}$Cl and $^{29,30}$Ar have been theoretically substantiated, and analogous deviations predicted for the lighter chlorine and argon isotopes. The  limits of nuclear structure existence are predicted for Ar and Cl isotopic chains, with $^{26}$Ar and $^{25}$Cl found to be the lightest sufficiently long-living nuclear systems. By simultaneous measurements of protons and $\gamma$-rays following decays of such systems as well as their $\beta$-delayed emission,  an interesting synergy effect may be achieved, which is demonstrated by the example of $^{30}$Cl and $^{31}$Ar ground state studies. Such synergy effect may be provided by the new EXPERT setup (EXotic Particle Emission and Radioactivity by Tracking), being operated inside the fragment separator and spectrometer facility at GSI, Darmstadt.
\end{abstract}


\maketitle


\section{Introduction}


Several states in proton ($p$) unbound isotopes $^{28}$Cl, $^{30}$Cl and $^{29}$Ar were reported recently  \cite{Mukha:2018}. This work continues the research published in Refs.\ \cite{Mukha:2015,Golubkova:2016,Xu:2018,Mukha:2018}. The systematics and cluster model studies in \cite{Mukha:2018} allowed to interpret the data as observations of ground state (g.s.) in $^{28}$Cl, g.s.\ and three excited states in $^{30}$Cl, and one state in $^{29}$Ar (either ground or excited state). Also the reported spectrum of $^{31}$Ar allowed for prescription of the g.s.\ energy of this isotope by using the isobaric symmetry systematics. Together with the known $p$-unbound isotopes $^{14,15,16}$F, the studied argon and chlorine isotopes constitute the most deeply-studied particle-unstable isotopic chains in the whole $Z\leq 20$ nuclei region.

In this work we continue the ``excursion beyond the proton dripline'' of Ref.\ \cite{Mukha:2018}. We intend to answer the question: What impact the obtained experimental results may have on our understanding of prospects to study the other nuclides located far (e.g., 2--5 mass units) beyond the driplines? Correspondingly, we discuss three main topics:

(i) The previously-published systematics of one-proton ($1p$) separation energies \cite{Mukha:2018} are extrapolated further into the unexplored region beyond the proton dripline.  The obtained results for the experimentally observed cases ($^{28-30}$Cl nuclides) are considerably different from the systematic trends available in the literature \cite{www:nndc2,Audi:2014,Tian:2013}. We extrapolate this systematics to the lightest chlorine and argon isotopes in Section \ref{sec:system}. The smaller than expected values of decay energies suggest longer-living states, and, consequently, weaker limitations on the nuclear structure existence beyond the dripline.

(ii)  We clarify the prospects of a limit of the nuclear structure existence by using the obtained information on the separation energies. We assume that a nuclear configuration has an individual structure with at least one distinctive state, if the orbiting valence protons of the system are reflected from the corresponding nuclear barrier at least one time. Thus nuclear half-life may be used as a gauge of such a limit. It is clear that the very long-lived particle-emitting states are \emph{quasistationary}. This means that they can be considered as \emph{stationary for majority of practical applications}. For example, the half-lives of all known heavy two-proton ($2p$) radioactivity cases ($^{45}$Fe, $^{48}$Ni, and $^{54}$Zn) are of few seconds. Thus, their $2p$ decays are so slow that weak transitions become their competitors with branching ratii of dozens of percent \cite{Pfutzner:2012}. We may assume that  modification of nuclear structure by continuum coupling is absolutely negligible for such states. In contrast, the continuum coupling becomes increasingly important for broad ground states beyond the driplines. For example, see the discussion connected with studies of the $^{10}$He g.s.\ in Ref.\ \cite{Sharov:2014}. This work demonstrated that the observed continuum properties of $^{10}$He can be crucially modified by peculiarities of initial nuclear structure of the reaction participants for the widespread experimental approaches (e.g.\ knockout reactions). Such a situation can be regarded as transitional to \emph{continuum dynamics}, where observable continuum response is also defined by the reaction mechanism and initial nuclear structure. Here the properties, interpretable as nuclear structure of the reaction products, cannot be reliably extracted from measured data. For example, we may refer to the well-known tetra-neutron system in continuum \cite{Grigorenko:2004}, where such an ambiguity has been demonstrated by applying the realistic scenario of the tetra-neutron population. Within the topic of the above discussion, we predict the limits of nuclear structure existence to be near the $^{25}$Cl and $^{26}$Ar isotopes in Section \ref{sec:limits}.

\begin{figure}
\begin{center}
\includegraphics[width=0.485\textwidth]{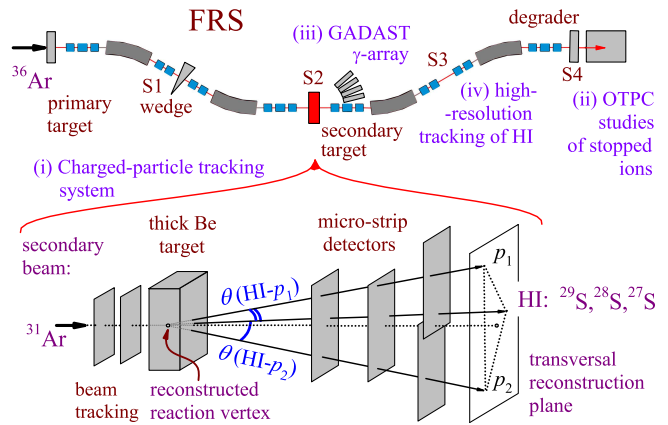}
\end{center}
\caption{The EXPERT pilot setup installed in the FRS fragment separator. (i) Charged particle tracking system shown in the lower inset consisted of beam tracking Si detectors providing energy loss and timing information and micro-strip Si detectors for precise tracking of the decay products of nuclei of interest. (ii) Optical Time projection Chamber (OTPC) for detection of radioactivity in the millisecond range. (iii) Array of $\gamma$-ray detectors around secondary target, GADAST. (iv) Detectors for identification of heavy ions and precise measurements of their momenta.}
\label{fig:layout}
\end{figure}

(iii) The experimental setup, used in Refs.\ \cite{Mukha:2015,Golubkova:2016,Xu:2018,Mukha:2018}, is a pilot version of the EXPERT (EXotic Particle Emission and Radioactivity by Tracking) setup planned by the Super-FRS Experiment Collaboration of the FAIR project, see Refs.\  \cite{www:expert-tdr,Aysto:2016} and Fig.\ \ref{fig:layout}. The tracking system for light ions and $\gamma$-ray detector were installed downstream of the secondary target in the internal focal plane of the fragment separator FRS at GSI, Darmstadt (see the details in Ref.\ \cite{Mukha:2018}). The first half of FRS was set for production and separation of $^{31}$Ar ions, and the second half was used as a spectrometer for heavy-ion decay products. The optical time projection chamber (OTPC) installed at S4 was studying beta-delayed particle emission and radioactive decays of heavy fragments living long enough to pass through the 30 m of the S2--S4 second half of the FRS. In this paper we demonstrate that the complementary measurements performed by all components of the EXPERT setup can be combined together, which allows for synergy effect in studies of the above-mentioned unbound nuclear systems. Such an effect is demonstrated in Section \ref{sec:synergy} by examples of $^{30}$Cl and $^{31}$Ar studies.

\section{The theoretical models applied}
\label{sec:th}


We apply several simple theoretical tools in this work.

The systematics of 1\emph{p} separation energies $S_p$ in chlorine isotopes are studied by using the  core+$p$ potential cluster model from \cite{Mukha:2018}. The systematic of Coulomb displacement energies in such a model is sensitive to two basic parameters: (i) an orbital size (governed by a potential radius) and (ii) charge radius of the core. These parameters are varied in the model in a systematic way.

The systematics of 2\emph{p} separation energies $S_{2p}$ in argon isotopes are based on (i) the $S_p$ values obtained from systematics of the related chlorine isotopes, and (ii) systematics of odd-even staggering energies based on the corresponding long isotopic and isotonic chains. This approach was actively used in our previous works \cite{Mukha:2015,Mukha:2018}, and it has proven to be very reliable tool with easily-estimated uncertainties.

The 1$p$-decay widths of chlorine isotopes are calculated by using the above-mentioned potential cluster model. We assume that the internal normalization of continuum states is an indicator of their resonance behavior. Such an indicator is more tractable for the broad states in comparison with the corresponding behavior of the phase shifts. We also use the R-matrix model for the 1$p$-decay width estimates in case of very narrow states.

The three-body $2p$-decay widths of argon isotopes are estimated by using the R-matrix-type model from Ref.\ \cite{Pfutzner:2012}. This model can be traced back to the three-body approximation with a simplified three-body Hamiltonian, which neglects nucleon-nucleon interaction \cite{Grigorenko:2007,Grigorenko:2007a}. The widths calculated by this model match the corresponding calculations of the complete three-body model within the factor of ten in the worst case. In the specific case of $2p$-decay width estimates of $^{26}$Ar, we use the sophisticated three-body core+$2p$ cluster model developed for its mirror isobaric partner $^{26}$O in \cite{Grigorenko:2013,Grigorenko:2015b}.

The unit system $\hbar=c=1$ is used in this work.


\section{Chlorine and Argon isotopic chains far beyond the proton dripline}
\label{sec:system}


The isotopes between $^{32}$Cl and $^{28}$Cl have been studied in Ref.\ \cite{Mukha:2018} by applying the two-body cluster $^A$S+$p$ model. The major parameters of the model (potential and charge radii of the sulphur core nucleus) were systematically varied (see Table I in \cite{Mukha:2018}). The Thomas-Ehrman effect \cite{Ehrman:1951,Thomas:1952}, especially pronounced in the $s$-$d$ shell nuclei is well accounted by such a model. As a result, the consistent description of the known low-lying spectra of $^{32}$Cl and $^{31}$Cl was obtained as well as the reasonable explanation of the newly observed states in $^{30}$Cl, $^{29}$Cl, and $^{28}$Cl nuclei.

\begin{figure}
\begin{center}
\includegraphics[width=0.485\textwidth]{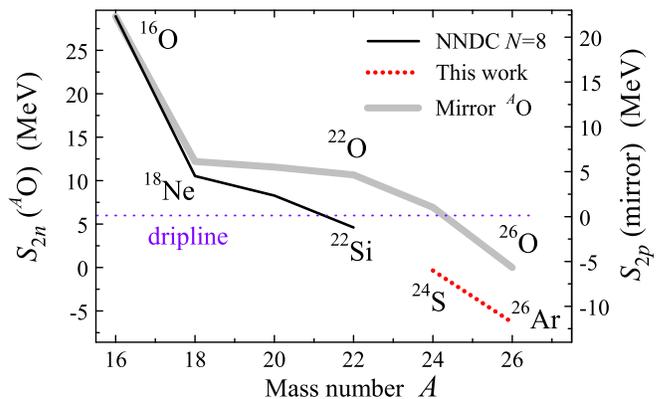}
\end{center}
\caption{The $S_{2p}$ estimates for $^{24}$S, see also Table \ref{tab:sep-s}. Two-neutron separation energies $S_{2n}$ for oxygen isotopic chain from \cite{www:nndc2} are shown by the thick gray line, two-proton separation energies $S_{2p}$ for the mirror isotope chains are shown by the solid black line. Red dotted line corresponds to the calculated $S_{2p}$ value for $^{26}$Ar (see Sec.\ \ref{sec:limits} and Fig.\ \ref{fig:continuum-cl-ar}) and the linear interpolation for $^{24}$S.}
\label{fig:o-s2n}
\end{figure}

Here we estimate the further isotopes beyond the proton dripline: $^{25-27}$Cl and $^{26-28}$Ar. The problem here is that for the lighter chlorine isotopes, the respective ``core nuclei'' $^{24-26}$S are particle-unbound with separation energies estimated in Table \ref{tab:sep-s}. These estimates are partly illustrated in Figure \ref{fig:o-s2n}. So, the main decay channels are expected to be $2p$, $3p$, and $4p$ emission from $^{26}$S, $^{25}$S, and $^{24}$S, respectively. One may notice that the decay energies of various decay branches of sulphur isotopes are much smaller than those of $1p$ emission from chlorine or $2p$ emission from argon isotopes. This means that the decay mechanism of $^{25-27}$Cl should be sequential emission of one proton followed by emission of $2-4$ protons from the respective sulphur daughter. Similarly, the decay mechanism of $^{26-28}$Ar should be sequential emission of two protons followed by the emission of $2-4$ protons. The half-life values of such sequential decays are practically entirely defined by the first ``fast'' step of sequential proton emission with large $Q_{2p}$ value. Therefore we will not take into account particle-instability of $^{24-26}$S in the following half-life estimates.

\begin{table}[b]
\caption{Estimated two-proton $S_{2p}$, three-proton $S_{3p}$, and four-proton $S_{4p}$ separation energies in MeV for three sulphur isotopes beyond the proton dripline.}
\begin{ruledtabular}
\begin{tabular}[c]{cccc}
Isotope & $S_{2p}$ & $S_{3p}$ & $S_{4p}$  \\
\hline
$^{26}$S  &  $-1.3$  &  2.0     & 2.1     \\
$^{25}$S  &  $-3.0$  &  $-5.3$  & $-3.5$  \\
$^{24}$S  &  $-6.0$  &  $-8.1$  & $-5.4$  \\
\end{tabular}
\end{ruledtabular}
\label{tab:sep-s}
\end{table}

\begin{figure}
\begin{center}
\includegraphics[width=0.48\textwidth]{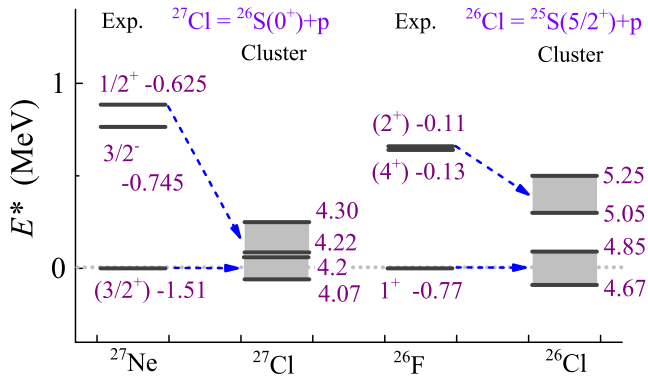}
\end{center}
\caption{Energy levels of $^{26}$Cl and $^{27}$Cl isotopes compared with their  mirror levels in isobaric partners $^{26}$F and $^{27}$Ne. Vertical axis shows excitation energies $E^*$. The legends for levels give spin-parity $J^{\pi}$ and energies relative to the $1p$-emission threshold for the Cl chain members or $1n$-emission threshold for their isobaric mirror partners. The given uncertainty of the states is due to variation of unknown charge radii of unstable sulphur daughter nuclei, see Fig.\ 12 in \cite{Mukha:2018}.}
\label{fig:levels-27cl-26c}
\end{figure}

The results of the cluster $^A$S+$p$ model calculations from Ref.\ \cite{Mukha:2018} for $^{26}$Cl and $^{27}$Cl are shown in Figure \ref{fig:levels-27cl-26c}. For calculation of $^{25}$Cl we used the $^{24}$O+$n$ potential developed for studies of the $^{26}$O in Ref.\cite{Grigorenko:2015b}. The $^{25}$O spectrum is quite ``poor'': it contains just one known $d$-wave $3/2^+$ state \cite{Hoffman:2008,Kondo:2016,Jones:2017}. By adding Coulomb interaction to the potential we obtain the $^{25}$Cl g.s.\ at $E_r = -S_p = 6.0-6.3$ MeV. The $E_r$ uncertainty here is defined by the $^{24}$S ``charge radius'' uncertainty taken in accordance with Fig.\ 12 of \cite{Mukha:2018}.

\begin{figure}
\begin{center}
\includegraphics[width=0.485\textwidth]{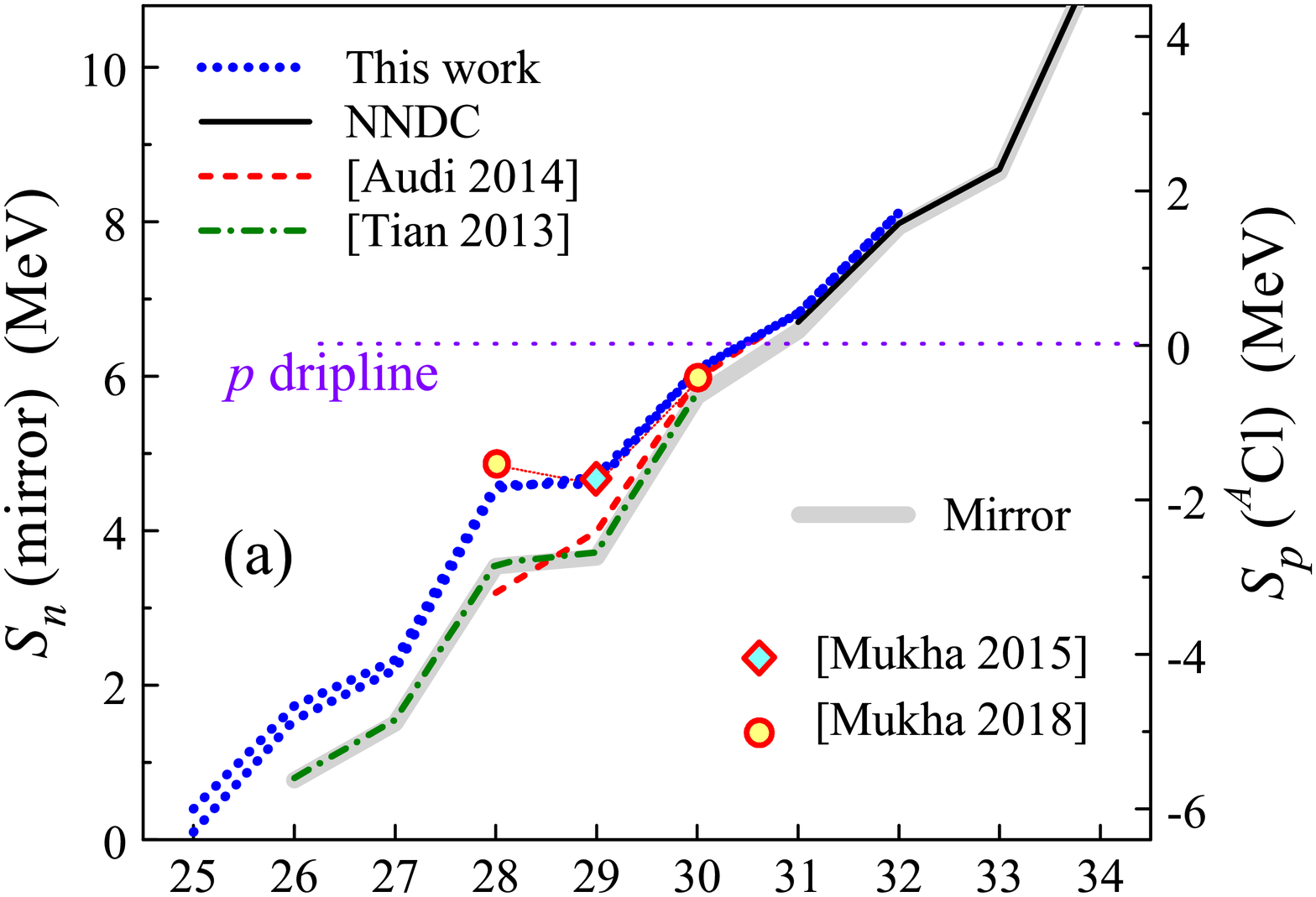}
\includegraphics[width=0.487\textwidth]{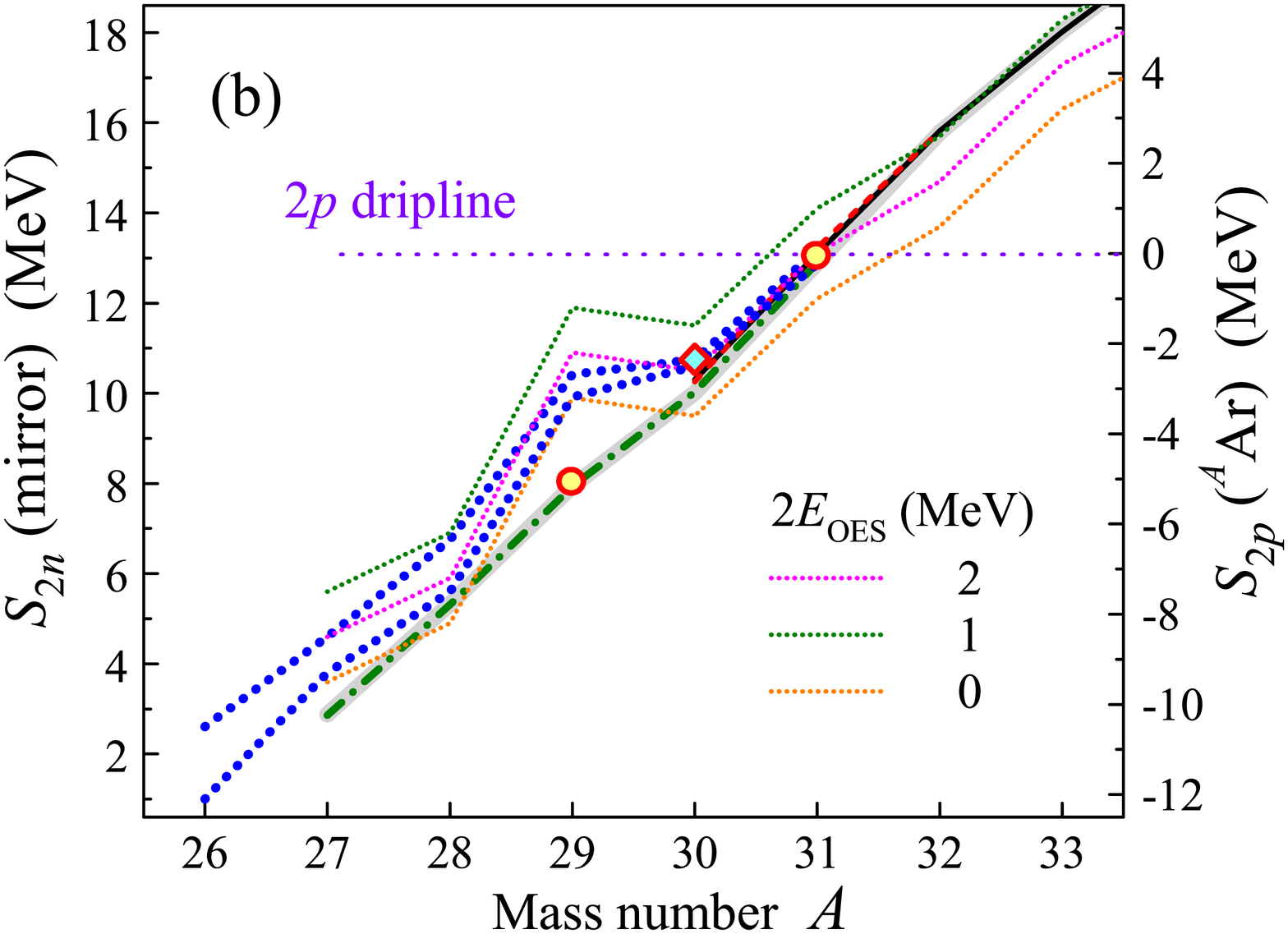}
\end{center}
\caption{Separation energies $S_{p}$ for the chlorine isotopic chain (a) and $S_{2p}$ for argon one (b) from \cite{www:nndc2} are shown by the solid black lines. Separation energies $S_{n}$ and $S_{2n}$ for the mirror isotone chains are shown by the thick gray lines. The systematic evaluations from \cite{Audi:2014} and \cite{Tian:2013} are given by the red dashed and green dash-dotted lines. The results of this work and \cite{Mukha:2018} based on cluster model and $E_{\text{OES}}$ systematics are shown by the blue dotted lines (there are two lines for upper and lower limiting estimates). The experimental values for $^{29}$Cl and $^{30}$Ar \cite{Mukha:2015} are shown by the red diamonds, while the results of \cite{Mukha:2018} are given by the red circles.}
\label{fig:sp-s2p}
\end{figure}

The systematics of proton separation energies $S_p$ for the chlorine isotopic chain is given in Figure \ref{fig:sp-s2p} (a). For illustration purpose we use here the data compiled in the NNDC database \cite{www:nndc2}, the standard AME2012  evaluation \cite{Audi:2014}, and the recent isobaric multiplet mass evaluation  \cite{Tian:2013}. One may see that the predicted systematics of \cite{Tian:2013} along the isobaric chain exactly follows the experimentally known systematics along the isotonic chain and can be regarded as trivial, while the predictions of \cite{Audi:2014} somewhat deviate from the isotone evolution. The predictions of our cluster model here and in \cite{Mukha:2018} (where they are supported by the data, see Table \ref{tab:sp-s2p}) demonstrate considerable deviations from the isotone expectation. These deviations have one major source --- the Thomas-Ehrman shift (TES) effect --- which is a well-established phenomenon and which is reliably described by the cluster model used in \cite{Mukha:2018} and here.

\begin{figure}
\begin{center}
\includegraphics[width=0.47\textwidth]{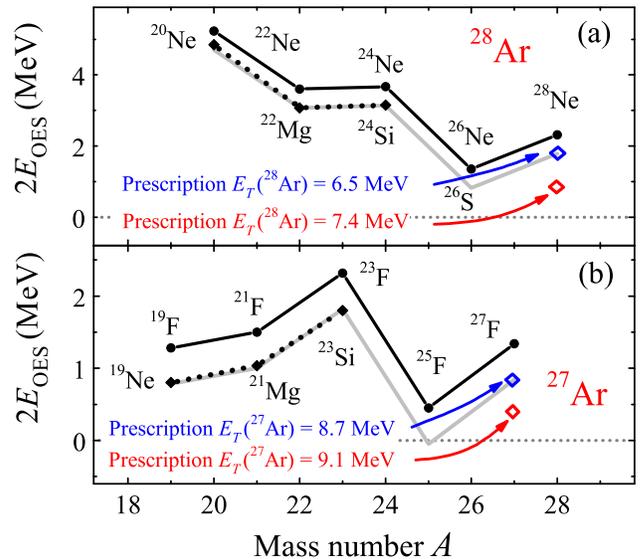}
\end{center}
\caption{Odd-even staggering energies $2E_{\text{OES}}=2S_N-S_{2N}$  for the isotones leading to $^{28}$Ar (a) and $^{27}$Ar (b) are shown by dotted line. The OES energies for the mirror isobar are given by solid line. Gray line is provided to guide the eye: this is solid line shifted down by constant values of about 0.5 MeV. The blue and red diamonds correspond to certain prescriptions of two-proton decay energy $E_T$ indicated in legends and giving odd-even staggering energies equal either its systematic value or half of this value.}
\label{fig:oes}
\end{figure}

On the basis of the developed $S_p$ systematics  for the chlorine isotopic chain, we can turn to the systematics studies of the argon isotopic chain. Following the approach of Ref.\ \cite{Mukha:2018} we apply the systematics of odd-even staggering energies (OES)
\[
2E_{\text{OES}} = S_{2p} - 2S_p \,,
\]
see Figure \ref{fig:oes}. For the chlorine isotopic chain beyond the dripline there is the trend of overbinding because of TES. For the argon isotopic chain there should be a competition of two trends: (i) overbinding because of TES (Coulomb displacement energy decreases because of an increase of the valence orbital size) and (ii) underbinding due to $E_{\text{OES}}$ reductions (pairing energy decreases because of an increase of the valence orbital size). This effect has been already emphasized in Ref.\ \cite{Mukha:2015}. Thus for the limiting estimates of the $S_{2p}$ in the argon isotopic chain we use the upper and lower estimates of $S_{p}$ shown in Fig.\ \ref{fig:sp-s2p} (a), which are then subtracted from the full $2E_{\text{OES}}$ value and multiplied by the factor of 1/2. The obtained results are shown in Fig.\ \ref{fig:sp-s2p} (b). The $S_p$ and $S_{2p}$ predicted for Ar and Cl isotopic chains are also collected in Table \ref{tab:sp-s2p}.

\begin{table}[b]
	\caption{The separation energies $S_{p}$ and $S_{2p}$ for chlorine and argon isotopes with mass \emph{ A } located beyond the proton dripline. The theoretical values are in the columns "theory", and the measured values are in the columns "exp., [Ref]" with the respective referencies. }
	\begin{ruledtabular}
		\begin{tabular}[c]{crrrr}
			& \multicolumn{2}{c}{$S_p(^{A}$Cl), MeV}  & \multicolumn{2}{c}{$S_{2p}(^{A}$Ar), MeV} \\
			$A$             &  theory$\quad$  & exp., [Ref]$\quad$ &  theory$\quad$ & exp., [Ref]$\quad$  \\
			\hline
			31  &             &    & $-0.08(15)$  &  0.006(34),  \cite{Mukha:2018}    \\
			30  &  $-0.311(1)$  & $-0.48(2)$, \cite{Mukha:2018}  & $-2.43(17)$  & $-2.45^{+5}_{-10}$, \cite{Golubkova:2016}      \\
			29  & $-1.75(1)$ & $-1.8(1)$, \cite{Mukha:2015} & $-2.93(25)$  &    $-5.50(18)$\footnotemark[1], \cite{Mukha:2018} \\
			28  &  $-1.83(2)$ & $-1.60(8)$, \cite{Mukha:2018}  & $-6.90(35)$  &       \\
			27  &  $-4.14(7)$ &    & $-8.90(40)$    &       \\
			26  &  $-4.66(9)$ &    & $-11.3(8)$  &       \\
			26  &   &    & $-11.7(3)$\footnotemark[2]  &       \\
			25  & $-6.15(15)$ &    &    &      \\
		\end{tabular}
	\end{ruledtabular}
	\label{tab:sp-s2p}
	\footnotetext[1]{Not clear either this is ground or excited state.}
	\footnotetext[2]{This theoretical result is obtained with the three-body model, see Fig.\ \ref{fig:continuum-cl-ar}.}
\end{table}

To conclude this section, the smaller than conventionally-expected  separation energies $S_{p}$ and $S_{2p}$ are predicted in this work for the chlorine and argon isotopes located far beyond the proton dripline. Such a general decrease should result in longer lifetimes of their ground and low-lying excited states, and consequently it may affect limits of existence of nuclear structure beyond the proton dripline.


\section{Limits of nuclear structure existence for Chlorine and Argon isotopic chains}
\label{sec:limits}


One of the fundamental tasks of nuclear science studies is determination of the limits of existence of individual states in nuclear systems. The half-life value can be chosen as a quantitative criterion of the nuclear structure formation. Let us consider a system formed by a potential barrier and assume that in order to form a nuclear state, there should be at least one reflection of the valence nucleon from the barrier. Then the potentials of $^AS$+$p$ channel used in \cite{Mukha:2018} and this work may help in estimations of such a limit for the chlorine isotopes by using the classical oscillation frequency
\[
\nu = \left( 2 \int_{r_1}^{r_2} \frac{dr}{v(r)} \right)^{-1} =
\left( \int_{r_1}^{r_2} dr\, \sqrt{\frac{2M}{E-V(r)}} \right)^{-1}\,,
\]
where $r_1$ and $r_2$ are two inner classical turning points. This oscillation frequency provide quite precise results (for sufficiently high barriers) when entering expression for quasiclassical estimate of width
\[
\Gamma = \nu \, P \,, \qquad P =  \int_{r_2}^{r_3} dr \, p(r)\,.
\]
For energies $E$ varying from 0 to $\sim 90\%$ of the barrier height, the estimate is $\nu \approx 1-3$ MeV. Thus we can assume that the systems with widths exceeding $3-5$ MeV have shorter half-lives than those needed for formation of the nuclear state. Such a system decays instantaneously, as, with a large probability, there will not be a single reflection from the potential barrier.

\begin{figure}
\begin{center}
\includegraphics[width=0.47\textwidth]{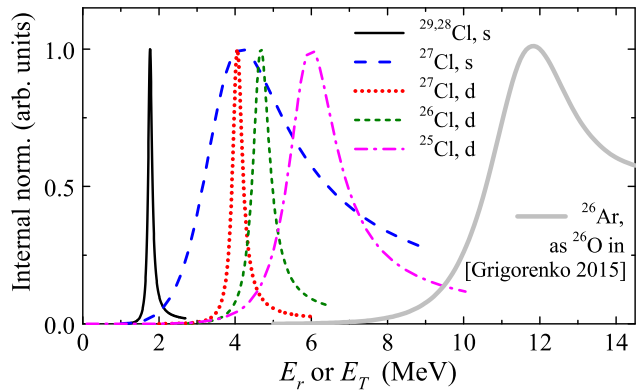}
\end{center}
\caption{Internal normalizations for the ground states of $^{25-29}$Cl isotopes as a function of proton decay energy $E_r$. The gray solid curve shows the excitation spectrum of $^{26}$Ar obtained in three-body model as a function of two-proton decay energy $E_T$. The same curve is given for both $^{28}$Cl and $^{29}$Cl, as the g.s.\ energies of these isotopes are predicted to be almost equal \cite{Mukha:2018}.}
\label{fig:continuum-cl-ar}
\end{figure}

The width values of the chlorine isotopes can be estimated from the calculated excitation spectra which are illustrated in Fig.\ \ref{fig:continuum-cl-ar}. For this purpose, we have used the internal normalization $N(E)$ of two-body continuum wave function $\psi(kr)$
\[
N(E) = \int_{0}^{r_2} dr \, |\psi(kr)|^2 \,,
\]
as a measure of the resonance formation. This is done in contrast to conventional scattering phase shifts which could not provide a firm signature of a resonance formation in the case of very  broad nuclear states ($\Gamma \gtrsim 1$ MeV). One may see in Fig.\ \ref{fig:continuum-cl-ar}, that $s$-wave states in chlorine isotopes become quite broad already in $^{27}$Cl ($\Gamma \gtrsim 3$ MeV). However, the $d$-wave states remain reasonably narrow  ($\Gamma \sim 1.5$ MeV) even in $^{25}$Cl with its quite high decay energy $E_r \sim 6 $ MeV.

\begin{figure}
\begin{center}
\includegraphics[width=0.48\textwidth]{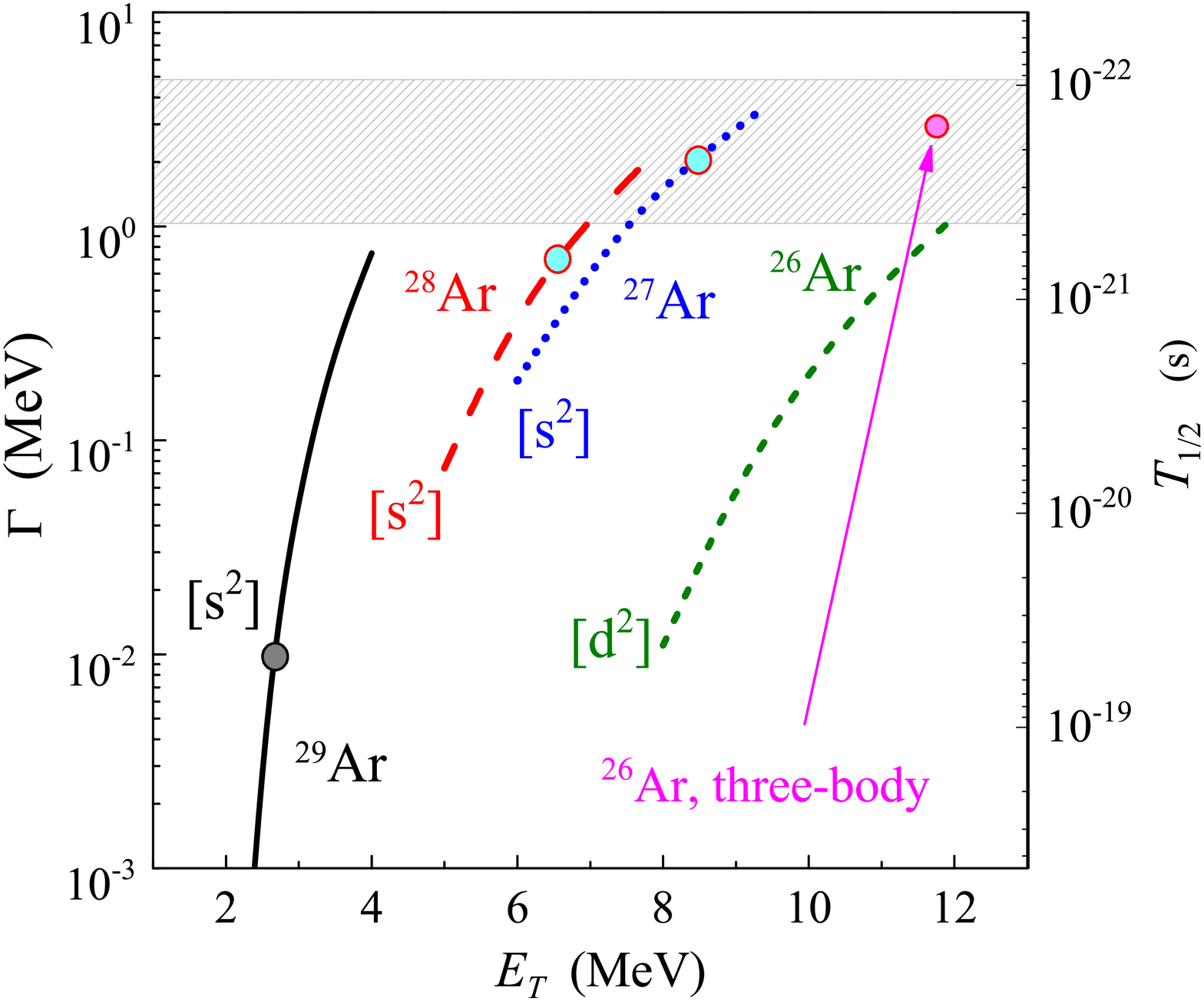}
\end{center}
\caption{Widths ${\Gamma}$ and half-lives $T_{1/2}$ of the $^{29}$Ar--$^{26}$Ar isotopes as functions of decay energy estimated by a direct decay model. The obtained decay energy of $^{29}$Ar \cite{Mukha:2018} is indicated by the black circle. The energies predicted in this work are indicated by the red-cyan circles. The magenta arrow points to the $\{E_T,\Gamma\}$ position evaluated for the $^{26}$Ar isotope by the three-body model, see Fig.\ \ref{fig:continuum-cl-ar}. The hatched area indicates the half-life range where the nuclear structure begins to ``dissolve''.}
\label{fig:lifetimes-ar}
\end{figure}

In Fig.\ \ref{fig:lifetimes-ar} we provide the \emph{upper-limit} width estimates for the Argon isotopes. They are performed in a ``direct decay'' R-matrix model from Ref.\ \cite{Pfutzner:2012}, where each proton is assumed to be in a resonant state of the core+$p$ subsystem with resonant energy $E_{j_i}$. The differential of the decay width is given by
\begin{eqnarray}
\frac{d\Gamma_{j_1j_2}(E_T)}{d \varepsilon } & = &
\frac{E_{T}\left \langle V_{3}\right \rangle^2}{2\pi} \, \frac{ \Gamma_{j_1} (\varepsilon E_{T})}{(\varepsilon E_{T}-E_{j_1})^{2}+\Gamma^{2}_{j_1}(\varepsilon E_{T})/4}
\nonumber \\
& \times & \frac{\Gamma_{j_2}((1-\varepsilon)E_{T})}
{((1-\varepsilon)E_{T}-E_{j_2})^{2} + \Gamma^{2}_{j_2}
((1-\varepsilon)E_{T})/4} , \quad \; \;
\label{eq:sequent}
\end{eqnarray}
where $j_i$ is the angular momentum of a core+$p_i$ subsystem. This model can be traced to the simplified Hamiltonian of the three-body system in which the nucleons interact with the core, but not with each other. The model approximates the true three-body decay mechanism and also provides a smooth transition to the sequential decay regime \cite{Grigorenko:2007,Grigorenko:2007a}. The matrix element $\left \langle V_{3}\right \rangle$ can be well approximated by
\[
\left \langle V_{3}\right \rangle ^{2}  = D_3 [(E_T-E_{j_1}-E_{j_2})^2 +
(\Gamma_{ph}(E_T))^2/4]\, ,
\]
where the parameter  $D_3 \approx 1.0-1.5$ (see Ref.\ \cite{Grigorenko:2007a}
for details), and $\Gamma_{ph}(E_T)$ should provide smooth width behavior around $E_T \sim E_{j_1}+E_{j_2}$. The assumed R-matrix parameters for the widths
\begin{equation}
\Gamma(E) = 2\, \frac{\theta^2}{2 M r^2_{\text{c}}} \, P_l(E, r_{\text{c}}, Z) \,,
\label{eq:r-matr-g}
\end{equation}
in the chlorine isotopes are given in Table \ref{tab:r-matr-param}. It was shown in \cite{Golubkova:2016} that the calculation has significant sensitivity only to the general decay parameters $\{E_T,E_r,\Gamma_r\}$.

\begin{table}[b]
\caption{The R-matrix parameters of the $^{A-2}$S+$p$ channel adopted by the width estimates of $^A$Ar isotopes: angular momentum $l$, the channel radius $r_{\text{c}}=1.2(A-1)^{1/3}$ in fm, the reduced width $\theta^2$, the resonance energy $E_r$ and corresponding width $\Gamma_r$ in MeV.}
\begin{ruledtabular}
\begin{tabular}[c]{cccccc}
$A$  & $l$  & $r_{\text{c}}$ & $\theta^2$ & $E_r$ & $\Gamma_r$  \\
\hline
26  & 2 & 3.31 & 1.0 & 6.0 & 0.5  \\
27  & 0 & 3.55 & 1.5 & 5.1 & 3.3  \\
28  & 0 & 3.60 & 1.5 & 4.2 & 2.2  \\
29  & 0 & 3.64 & 1.5 & 1.6 & $5.7 \times 10^{-3}$  \\
31  & 0 & 3.73 & 1.5 & 0.5 & $5.3 \times 10^{-6}$  \\
31  & 2 & 3.73 & 1.0 & 0.5 & $3.6 \times 10^{-8}$  \\
\end{tabular}
\end{ruledtabular}
\label{tab:r-matr-param}
\end{table}

For the width estimates presented in Fig.\ \ref{fig:lifetimes-ar}, we consider initial structure and decay of the Argon isotopes via $[s^2]_0$ configurations with $s$-wave resonance parameters inherited from the two-body model calculations for the chlorine isotopic chain. Such an assumption guarantees the upper-limit width estimate (see discussions in \cite{Pfutzner:2012,Grigorenko:2007,Grigorenko:2007a}). However, this does not work for $^{26}$Ar. The $^{25}$Cl isotope which is core+$p$ subsystem of $^{26}$Ar has very ``poor'' spectrum with just one low-energy $d$-wave state. For that reason we make a $[d^2]_0$ estimate for $^{26}$Ar decay. In order to cross check it, we made three-body calculations of excitation function in a full three-body model. It is known that for $2N$ decays of higher orbital configurations accounting for $N$-$N$ final state interaction may lead to a drastic decrease of the half-life \cite{Grigorenko:2013}. The three-body calculations are completely analogous to the calculations of $^{26}$O g.s.\ in Ref.\ \cite{Grigorenko:2015b} with the added Coulomb interaction in the $p$-$p$ and core-$p$ channels. The corresponding excitation function is shown in Fig.\ \ref{fig:continuum-cl-ar}. The obtained resonance energy $E_T \sim 11.7$ MeV is in a good agreement with systematic results of this work, see Fig.\ \ref{fig:sp-s2p} (b) and Table \ref{tab:sp-s2p}. The estimated width value $\Gamma \sim 3$ MeV is shown in Figure \ref{fig:lifetimes-ar}.

To conclude this Section, a number of relatively narrow states, which presumably can be interpreted in terms of nuclear structure, is predicted in the chlorine and argon isotopic chains down to $^{26}$Ar and $^{25}$Cl isotopes. These are located on $N=8$ shell closure and the lighter systems along these chains are not expected to exist. Population of such exotic systems is far beyond the reach of any modern experiment. However, we emphasize that there exists a rich, often not considered, research field far beyond the proton dripline which does not seem to be exhausted in the observable future.


\section{Synergy effect in the EXPERT setup}
\label{sec:synergy}


The experimental setup used in the works \cite{Mukha:2015,Golubkova:2016,Xu:2018,Mukha:2018} is a pilot version of the EXPERT (EXotic Particle Emission and Radioactivity by Tracking) project proposed for the physics program of the Super-FRS Experimental collaboration of the FAIR facility, see Refs.\ \cite{www:expert-tdr,Aysto:2016}. The EXPERT setup will be located mainly in the middle of the Super-FRS fragment separator which first part will produce and separate ions of interest, and the second part will measure momenta of heavy-ion decay products with high precision. The  EXPERT setup is being tested at the FRS fragment separator at GSI (Darmstadt). It consists of the following devices, see Fig.\ \ref{fig:layout}:  (i) charged-particle tracking system based on microstrip silicon detectors ($\mu$SSD) located downstream of the secondary target in the S2 middle focal plane of FRS, (ii) Optical time projection chamber (OTPC) at the end of FRS, (iii) $\gamma$-ray detectors around the secondary target GADAST. Important part of the EXPERT initiative is (iv) the use of the second half of FRS as a high-resolution spectrometer. This feature provides unique $\{A,Z\}$ identifications for a number of possible long-lived (i.e., with $T_{1/2}\gtrsim 100$ ns) heavy-ion reaction products and their implantation into the OTPC for radioactivity studies.

\begin{figure}
\begin{center}
\includegraphics[width=0.49\textwidth]{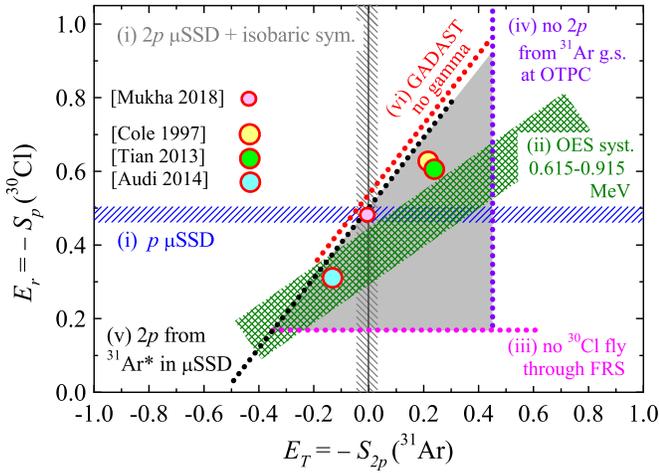}
\end{center}
\caption{The limitations on the correlated values of $S_p$ in $^{30}$Cl and $S_{2p}$ in $^{31}$Ar from different types of data and estimates, see text. The predictions of systematics studies \cite{Cole:1996,Audi:2014} are shown by circles.}
\label{fig:triangle}
\end{figure}

The instruments (i)--(iii) can be operated as independent devices and each of them has scientific value of their own. However, for studies of nuclear systems beyond the dripline, the elements of EXPERT  operated together provide an important synergy effect which has not been discussed so far. Let us demonstrate such a synergy effect by example of the $^{30}$Cl and $^{31}$Ar g.s.\ studies.

\begin{figure}
\begin{center}
\includegraphics[width=0.49\textwidth]{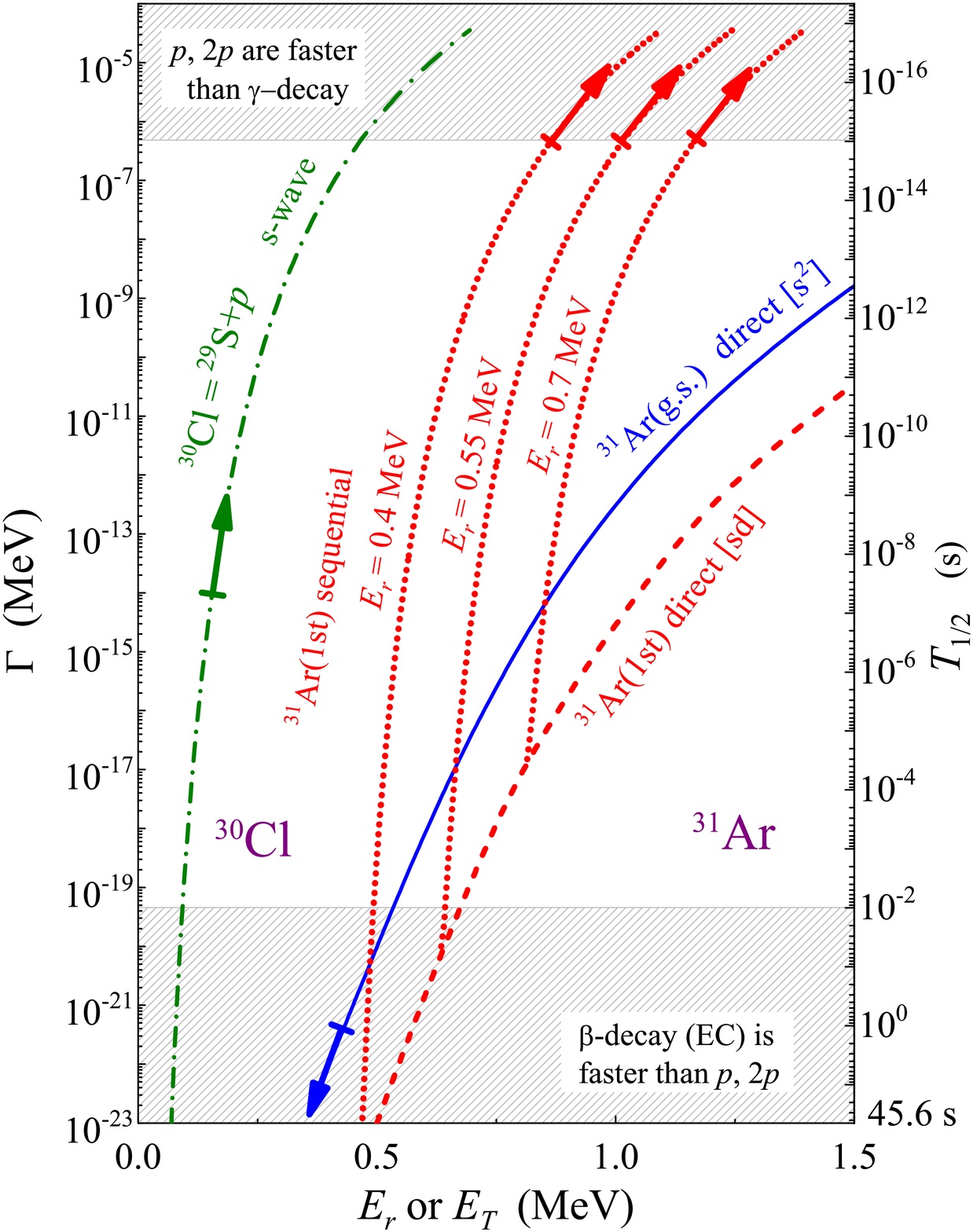}
\end{center}
\caption{Proton and two-proton decay widths ${\Gamma}$ and half-lives $T_{1/2}$ of $^{30}$Cl and $^{31}$Ar as a function of decay energies $E_r$ (for $p$-emission)  or $E_T$ (for $2p$-emission). True $2p$ decay of the $^{31}$Ar g.s.\ is shown by the solid blue curve. True $2p$ decay of the $^{31}$Ar first excited state is shown by the dashed red curve. Transition to sequential decay of $^{31}$Ar first excited state is illustrated by the dotted red curves for different $^{30}$Cl g.s. positions. The $1p$ decay of $^{30}$Cl g.s.\ (assuming $s$-wave emission) is shown by the green dash-dotted curve.}
\label{fig:lifetimes-cl-ar}
\end{figure}

Figure \ref{fig:triangle} shows the constrains that can be imposed on the ground state energies of $^{30}$Cl and $^{31}$Ar connected with different types of measurements and theoretical considerations given below. They are partly based on the half-life estimates of these isotopes presented in Fig.\ \ref{fig:lifetimes-cl-ar}. First, let us explain the Fig.\ \ref{fig:lifetimes-cl-ar}. The half-life of $^{30}$Cl is calculated for $^{29}$S+$p$ $s$-wave decay by the R-matrix model,see Eq.\ (\ref{eq:r-matr-g}). The half-life of $^{31}$Ar ground and first excited states are estimated by the R-matrix-type direct decay three-body model, see Eq.\ (\ref{eq:sequent}), Table \ref{tab:r-matr-param} and the corresponding discussion. The calculations are performed assuming the $[s^2]$ and $[sd]$ configurations in the $^{29}$S+$p$+$p$ channel, respectively. For the $^{31}$Ar first excited state the $2p$ decay energy $E_T \sim 1$ MeV is expected, while for the $^{30}$Cl g.s.\ the expectation is $E_r \sim 0.5$ MeV \cite{Mukha:2018}. Therefore for this state the turnover from a true $2p$ to a sequential $2p$ decay mechanism is expected at $E_T \gtrsim E_r$. These decay modes are characterized by very different behavior of width as a function of energy. We have estimated three half-life curves for the $^{31}$Ar first excited state corresponding to the  assumed $^{30}$Cl g.s.\ energies of 0.4, 0.55, 0.7 MeV, which are shown in Fig.\ \ref{fig:lifetimes-cl-ar} by the red doted curves.

One should note that the widths of states are estimated for the fastest possible $s$-wave proton emission from $^{30}$Cl as well as the fastest $[s^2]$-wave $2p$ decay from $^{31}$Ar g.s. We have also assumed that the first process in the decay of the $^{31}$Ar excited state is the emission of the $s$-wave proton, which is a very conservative estimate because the $^{30}$Cl g.s.\ has presumably an $s$-wave configuration. So, the more realistic half-life limitations could be even more stringent than those provided below.

Now we turn to a description of the obtained limits on decay energies of $^{31}$Ar and $^{30}$Cl, which are illustrated in Figure \ref{fig:triangle}. There are in total 6 different limitations connected with observation or/and non-observation of different states and decay channels in these systems.

\noindent (i) The horizontal and vertical hatched bands correspond to the energies directly inferred from the measurements by the $\mu$SSD tracking system as discussed above in this work and in Ref.\ \cite{Mukha:2018}.

\noindent (ii) The diagonal hatched band is provided on the basis of systematics of OES energies of Fig.\ 13 (a) from \cite{Mukha:2018}. We assume that isobaric symmetry for $^{31}$Ar is a good assumption giving $2E_{\text{OES}}=0.915$ MeV. In Fig.\ \ref{fig:triangle} we assume that some deviation from this value  ($-300$ keV) is possible but not too much, and $2E_{\text{OES}}=0.615$ MeV is taken as the lower limit.

\noindent (iii) The ions of $^{30}$Cl were not observed at the final focal plane of FRS. This means that the half-life of $^{30}$Cl is shorter than the  time-of-flight (ToF) through the S2--S4 section of FRS which is around 150 ns. We use the ToF value of 50 ns as the upper-limit estimate. This imposes the  corresponding lower-limit estimate $E_r> 160$ keV, see the green arrow in Fig.\ \ref{fig:lifetimes-cl-ar} and the magenta horizontal dotted line in Fig.\ \ref{fig:triangle}.

\noindent (iv) The $^{31}$Ar isotopes were implanted into the OTPC in order to study $\beta$-delayed proton emission \cite{Lis:2015}. A non-observation limit value is less than the obtained branching ratio of $7(2)\times 10^{-4}$ for the $\beta$-delayed decay channel of $^{31}$Ar. This means that the $^{31}$Ar g.s.\ energy is $E_T<0.4$ MeV, see the blue arrow in Fig.\ \ref{fig:lifetimes-cl-ar} and the vertical violet dotted line in Fig.\ \ref{fig:triangle}. Otherwise, the prompt $2p$ emission from $^{31}$Ar becomes faster than its $\beta$-decay.

\noindent (v) The estimated half-live curves for $2p$ decay of the $^{31}$Ar first excited state are given in Fig.\ \ref{fig:lifetimes-cl-ar}. It is clear that if the partial half-life of $^{31}$Ar with respect to $2p$ emission is longer than $\sim 1$ fs, then the preferable decay branch for this state is $\gamma$-deexcitation to the ground state. Since the $2p$ decay of the $^{31}$Ar first excited state is really observed, then the half-life limitations indicated by the red arrows in Fig.\ \ref{fig:lifetimes-cl-ar} infer synchronous limitations both on proton decay energy $E_r$ of $^{30}$Cl g.s.\ and two-proton decay energy $E_T$ of the $^{31}$Ar first excited state. The latter is transfered into $E_T$ of the $^{31}$Ar g.s.\ in Fig.\ \ref{fig:triangle} by subtracting 0.96 MeV as assumed from isobaric symmetry with $^{31}$Al in Ref.\ \cite{Mukha:2018} (inclined black dotted line). For example, let us consider the $E_r=0.7$ MeV curve in Fig.\ \ref{fig:lifetimes-cl-ar}. It provides the $E_T=1.21$ MeV limit and thus leads to the black dotted line passing through point $\{0.21,0.7\}$ in Fig.\ \ref{fig:triangle}.

\noindent (vi) Analogous information could be inferred from non-observation of $\gamma$-rays from the $\gamma$-decay of the $^{31}$Ar first excited state in GADAST (the inclined red line in Fig.\  \ref{fig:triangle}). The statistics in the current experiment was not sufficient to make this information significant, but in general case it could provide additional cross-check of consistency for the different types of the data.

All in all, the limitations shown in Fig.\ \ref{fig:triangle} lead together to a dramatic reduction of the area admissible for the correlated $^{30}$Cl vs.\ $^{31}$Ar g.s.\  energies compared to the data provided by the $\mu$SSD tracking detectors of the EXPERT only. We should state here that the confidence in the results for $^{30}$Cl and $^{31}$Ar g.s.\ energies is strongly enforced by the synergy analysis presented here.


\section{Summary}
%

In this work we use the data \cite{Mukha:2018} concerning the most remote from the proton dripline $^{30-28}$Cl and $^{31-29}$Ar isotopes, which allow for the further advances in studying an unknown domain beyond the proton dripline. The main results of this work are:

\noindent  (i) The systematic studies of the chlorine and argon isotopic chains beyond proton dripline have been performed. Large Thomas-Ehrmann shifts were revealed for the $^{29}$Cl and $^{30}$Ar isotopes in Ref.\ \cite{Mukha:2015}, and here we report further increased values in the $^{28}$Cl and $^{30}$Cl isotopes. The predictions for the very remote from the dripline isotopes $^{27}$Cl and $^{28}$Ar are provided by the elaborated models. For these isotopes, the Thomas-Ehrmann effect becomes less important as (a) the isobaric mirror partners of these nuclides are located in proximity of the neutron dripline and (b) the ground states are $d$-wave states which are less prone to modification by the Thomas-Ehrmann shift.

\noindent (ii) The obtained decay energies for the experimentally observed cases ($^{28-30}$Cl nuclides, \cite{Mukha:2018}) are considerably different (smaller) from the systematic trends available in the literature. The extrapolations to even lighter chlorine and argon isotopes also continue this trend. Smaller decay energy systematics means ``survival'' of the nuclear structure for even more remote from the dripline particle-unstable systems. The limits of nuclear structure existence for the proton-rich edge of chlorine and argon isotope chains are predicted to be in $^{26}$Ar and $^{25}$Cl.

\noindent (iii) Amazingly small $2p$-separation energy of $6(34)$ keV of the $^{31}$Ar ground state reported in the preceding article~\cite{Mukha:2018} has been explored in addition by using the complementary data available in the setup and relevant theoretical speculations. The synergy effect of the measurements performed by different detectors of the EXPERT setup was demonstrated. It gives the stronger support to the conclusions about the decays of $^{30}$Cl and $^{31}$Ar isotopes.


\begin{acknowledgments}
This work was supported in part by the Helmholtz International Center for FAIR (HIC for FAIR), the Helmholtz Association (grant IK-RU-002), the Russian Science Foundation (grant No.\ 17-12-01367), the Polish National Science Center (Contract No.\ UMO-2015/17/B/ST2/00581), the Polish Ministry of Science and Higher Education (Grant No.\ 0079/DIA/2014/43, Grant Diamentowy), the Helmholtz- CAS Joint Research Group (grant HCJRG-108), the FPA2016-77689-C2-1-R contract (MEC, Spain), the MEYS Projects  LTT17003 and LM2015049 (Czech Republic), the Justus-Liebig-Universit\"at Giessen (JLU) and GSI under the JLU-GSI strategic Helmholtz partnership agreement.
\end{acknowledgments}


\bibliographystyle{apsrev4-1}

\bibliography{/u/dkostyl/PRC_paper_Ar/references/all_theor}

\begin{thebibliography}{23}%
\makeatletter
\providecommand \@ifxundefined [1]{%
 \@ifx{#1\undefined}
}%
\providecommand \@ifnum [1]{%
 \ifnum #1\expandafter \@firstoftwo
 \else \expandafter \@secondoftwo
 \fi
}%
\providecommand \@ifx [1]{%
 \ifx #1\expandafter \@firstoftwo
 \else \expandafter \@secondoftwo
 \fi
}%
\providecommand \natexlab [1]{#1}%
\providecommand \enquote  [1]{``#1''}%
\providecommand \bibnamefont  [1]{#1}%
\providecommand \bibfnamefont [1]{#1}%
\providecommand \citenamefont [1]{#1}%
\providecommand \href@noop [0]{\@secondoftwo}%
\providecommand \href [0]{\begingroup \@sanitize@url \@href}%
\providecommand \@href[1]{\@@startlink{#1}\@@href}%
\providecommand \@@href[1]{\endgroup#1\@@endlink}%
\providecommand \@sanitize@url [0]{\catcode `\\12\catcode `\$12\catcode
  `\&12\catcode `\#12\catcode `\^12\catcode `\_12\catcode `\%12\relax}%
\providecommand \@@startlink[1]{}%
\providecommand \@@endlink[0]{}%
\providecommand \url  [0]{\begingroup\@sanitize@url \@url }%
\providecommand \@url [1]{\endgroup\@href {#1}{\urlprefix }}%
\providecommand \urlprefix  [0]{URL }%
\providecommand \Eprint [0]{\href }%
\providecommand \doibase [0]{http://dx.doi.org/}%
\providecommand \selectlanguage [0]{\@gobble}%
\providecommand \bibinfo  [0]{\@secondoftwo}%
\providecommand \bibfield  [0]{\@secondoftwo}%
\providecommand \translation [1]{[#1]}%
\providecommand \BibitemOpen [0]{}%
\providecommand \bibitemStop [0]{}%
\providecommand \bibitemNoStop [0]{.\EOS\space}%
\providecommand \EOS [0]{\spacefactor3000\relax}%
\providecommand \BibitemShut  [1]{\csname bibitem#1\endcsname}%
\let\auto@bib@innerbib\@empty
\bibitem [{\citenamefont {Mukha}\ \emph {et~al.}(2018)\citenamefont {Mukha},
  \citenamefont {Grigorenko}, \citenamefont {Kostyleva}, \citenamefont
  {Acosta}, \citenamefont {Casarejos}, \citenamefont {Dominik}, \citenamefont
  {Du\'{e}nas-D\'{i}az}, \citenamefont {Dunin}, \citenamefont {Espino},
  \citenamefont {Estrad\'{e}}, \citenamefont {Fari\'{n}on}, \citenamefont
  {Fomichev}, \citenamefont {Geissel}, \citenamefont {Gorshkov}, \citenamefont
  {Janas}, \citenamefont {Kami\'{n}ski}, \citenamefont {Kiselev}, \citenamefont
  {Kn\"{o}bel}, \citenamefont {Krupko}, \citenamefont {Kuich}, \citenamefont
  {Lis}, \citenamefont {Litvinov}, \citenamefont {Marquinez-Dur\'{a}n},
  \citenamefont {Martel}, \citenamefont {Mazzocchi}, \citenamefont {Nociforo},
  \citenamefont {Ord\'{u}z}, \citenamefont {Pf\"{u}tzner}, \citenamefont
  {Pietri}, \citenamefont {Pomorski}, \citenamefont {Prochazka}, \citenamefont
  {Rymzhanova}, \citenamefont {S\'{a}nchez-Ben\'{i}tez}, \citenamefont
  {Scheidenberger}, \citenamefont {Sharov}, \citenamefont {Simon},
  \citenamefont {Sitar}, \citenamefont {Slepnev}, \citenamefont {Stanoiu},
  \citenamefont {Strmen}, \citenamefont {Szarka}, \citenamefont {Takechi},
  \citenamefont {Tanaka}, \citenamefont {Weick}, \citenamefont {Winkler},
  \citenamefont {Winfield}, \citenamefont {Xu},\ and\ \citenamefont
  {Zhukov}}]{Mukha:2018}%
  \BibitemOpen
  \bibfield  {author} {\bibinfo {author} {\bibfnamefont {I.}~\bibnamefont
  {Mukha}}, \bibinfo {author} {\bibfnamefont {L.~V.}\ \bibnamefont
  {Grigorenko}}, \bibinfo {author} {\bibfnamefont {D.}~\bibnamefont
  {Kostyleva}}, \bibinfo {author} {\bibfnamefont {L.}~\bibnamefont {Acosta}},
  \bibinfo {author} {\bibfnamefont {E.}~\bibnamefont {Casarejos}}, \bibinfo
  {author} {\bibfnamefont {W.}~\bibnamefont {Dominik}}, \bibinfo {author}
  {\bibfnamefont {J.}~\bibnamefont {Du\'{e}nas-D\'{i}az}}, \bibinfo {author}
  {\bibfnamefont {V.}~\bibnamefont {Dunin}}, \bibinfo {author} {\bibfnamefont
  {J.~M.}\ \bibnamefont {Espino}}, \bibinfo {author} {\bibfnamefont
  {A.}~\bibnamefont {Estrad\'{e}}}, \bibinfo {author} {\bibfnamefont
  {F.}~\bibnamefont {Fari\'{n}on}}, \bibinfo {author} {\bibfnamefont
  {A.}~\bibnamefont {Fomichev}}, \bibinfo {author} {\bibfnamefont
  {H.}~\bibnamefont {Geissel}}, \bibinfo {author} {\bibfnamefont
  {A.}~\bibnamefont {Gorshkov}}, \bibinfo {author} {\bibfnamefont
  {Z.}~\bibnamefont {Janas}}, \bibinfo {author} {\bibfnamefont
  {G.}~\bibnamefont {Kami\'{n}ski}}, \bibinfo {author} {\bibfnamefont
  {O.}~\bibnamefont {Kiselev}}, \bibinfo {author} {\bibfnamefont
  {R.}~\bibnamefont {Kn\"{o}bel}}, \bibinfo {author} {\bibfnamefont
  {S.}~\bibnamefont {Krupko}}, \bibinfo {author} {\bibfnamefont
  {M.}~\bibnamefont {Kuich}}, \bibinfo {author} {\bibfnamefont {A.~A.}\
  \bibnamefont {Lis}}, \bibinfo {author} {\bibfnamefont {Y.~A.}\ \bibnamefont
  {Litvinov}}, \bibinfo {author} {\bibfnamefont {G.}~\bibnamefont
  {Marquinez-Dur\'{a}n}}, \bibinfo {author} {\bibfnamefont {I.}~\bibnamefont
  {Martel}}, \bibinfo {author} {\bibfnamefont {C.}~\bibnamefont {Mazzocchi}},
  \bibinfo {author} {\bibfnamefont {C.}~\bibnamefont {Nociforo}}, \bibinfo
  {author} {\bibfnamefont {A.~K.}\ \bibnamefont {Ord\'{u}z}}, \bibinfo {author}
  {\bibfnamefont {M.}~\bibnamefont {Pf\"{u}tzner}}, \bibinfo {author}
  {\bibfnamefont {S.}~\bibnamefont {Pietri}}, \bibinfo {author} {\bibfnamefont
  {M.}~\bibnamefont {Pomorski}}, \bibinfo {author} {\bibfnamefont
  {A.}~\bibnamefont {Prochazka}}, \bibinfo {author} {\bibfnamefont
  {S.}~\bibnamefont {Rymzhanova}}, \bibinfo {author} {\bibfnamefont {A.~M.}\
  \bibnamefont {S\'{a}nchez-Ben\'{i}tez}}, \bibinfo {author} {\bibfnamefont
  {C.}~\bibnamefont {Scheidenberger}}, \bibinfo {author} {\bibfnamefont
  {P.}~\bibnamefont {Sharov}}, \bibinfo {author} {\bibfnamefont
  {H.}~\bibnamefont {Simon}}, \bibinfo {author} {\bibfnamefont
  {B.}~\bibnamefont {Sitar}}, \bibinfo {author} {\bibfnamefont
  {R.}~\bibnamefont {Slepnev}}, \bibinfo {author} {\bibfnamefont
  {M.}~\bibnamefont {Stanoiu}}, \bibinfo {author} {\bibfnamefont
  {P.}~\bibnamefont {Strmen}}, \bibinfo {author} {\bibfnamefont
  {I.}~\bibnamefont {Szarka}}, \bibinfo {author} {\bibfnamefont
  {M.}~\bibnamefont {Takechi}}, \bibinfo {author} {\bibfnamefont {Y.~K.}\
  \bibnamefont {Tanaka}}, \bibinfo {author} {\bibfnamefont {H.}~\bibnamefont
  {Weick}}, \bibinfo {author} {\bibfnamefont {M.}~\bibnamefont {Winkler}},
  \bibinfo {author} {\bibfnamefont {J.~S.}\ \bibnamefont {Winfield}}, \bibinfo
  {author} {\bibfnamefont {X.}~\bibnamefont {Xu}}, \ and\ \bibinfo {author}
  {\bibfnamefont {M.~V.}\ \bibnamefont {Zhukov}},\ }\href@noop {} {\bibfield
  {journal} {\bibinfo  {journal} {arXiv:1803.10951, Submitted to Phys. Rev. C}\
  } (\bibinfo {year} {2018})}\BibitemShut {NoStop}%
\bibitem [{\citenamefont {Mukha}\ \emph {et~al.}(2015)\citenamefont {Mukha},
  \citenamefont {Grigorenko}, \citenamefont {Xu}, \citenamefont {Acosta},
  \citenamefont {Casarejos}, \citenamefont {Ciemny}, \citenamefont {Dominik},
  \citenamefont {Du\'{e}nas-D\'{i}az}, \citenamefont {Dunin}, \citenamefont
  {Espino}, \citenamefont {Estrad\'{e}}, \citenamefont {Fari\'{n}on},
  \citenamefont {Fomichev}, \citenamefont {Geissel}, \citenamefont {Golubkova},
  \citenamefont {Gorshkov}, \citenamefont {Janas}, \citenamefont
  {Kami\'{n}ski}, \citenamefont {Kiselev}, \citenamefont {Kn\"{o}bel},
  \citenamefont {Krupko}, \citenamefont {Kuich}, \citenamefont {Litvinov},
  \citenamefont {Marquinez-Dur\'{a}n}, \citenamefont {Martel}, \citenamefont
  {Mazzocchi}, \citenamefont {Nociforo}, \citenamefont {Ord\'{u}z},
  \citenamefont {Pf\"{u}tzner}, \citenamefont {Pietri}, \citenamefont
  {Pomorski}, \citenamefont {Prochazka}, \citenamefont {Rymzhanova},
  \citenamefont {S\'{a}nchez-Ben\'{i}tez}, \citenamefont {Scheidenberger},
  \citenamefont {Sharov}, \citenamefont {Simon}, \citenamefont {Sitar},
  \citenamefont {Slepnev}, \citenamefont {Stanoiu}, \citenamefont {Strmen},
  \citenamefont {Szarka}, \citenamefont {Takechi}, \citenamefont {Tanaka},
  \citenamefont {Weick}, \citenamefont {Winkler}, \citenamefont {Winfield},\
  and\ \citenamefont {Zhukov}}]{Mukha:2015}%
  \BibitemOpen
  \bibfield  {author} {\bibinfo {author} {\bibfnamefont {I.}~\bibnamefont
  {Mukha}}, \bibinfo {author} {\bibfnamefont {L.~V.}\ \bibnamefont
  {Grigorenko}}, \bibinfo {author} {\bibfnamefont {X.}~\bibnamefont {Xu}},
  \bibinfo {author} {\bibfnamefont {L.}~\bibnamefont {Acosta}}, \bibinfo
  {author} {\bibfnamefont {E.}~\bibnamefont {Casarejos}}, \bibinfo {author}
  {\bibfnamefont {A.~A.}\ \bibnamefont {Ciemny}}, \bibinfo {author}
  {\bibfnamefont {W.}~\bibnamefont {Dominik}}, \bibinfo {author} {\bibfnamefont
  {J.}~\bibnamefont {Du\'{e}nas-D\'{i}az}}, \bibinfo {author} {\bibfnamefont
  {V.}~\bibnamefont {Dunin}}, \bibinfo {author} {\bibfnamefont {J.~M.}\
  \bibnamefont {Espino}}, \bibinfo {author} {\bibfnamefont {A.}~\bibnamefont
  {Estrad\'{e}}}, \bibinfo {author} {\bibfnamefont {F.}~\bibnamefont
  {Fari\'{n}on}}, \bibinfo {author} {\bibfnamefont {A.}~\bibnamefont
  {Fomichev}}, \bibinfo {author} {\bibfnamefont {H.}~\bibnamefont {Geissel}},
  \bibinfo {author} {\bibfnamefont {T.~A.}\ \bibnamefont {Golubkova}}, \bibinfo
  {author} {\bibfnamefont {A.}~\bibnamefont {Gorshkov}}, \bibinfo {author}
  {\bibfnamefont {Z.}~\bibnamefont {Janas}}, \bibinfo {author} {\bibfnamefont
  {G.}~\bibnamefont {Kami\'{n}ski}}, \bibinfo {author} {\bibfnamefont
  {O.}~\bibnamefont {Kiselev}}, \bibinfo {author} {\bibfnamefont
  {R.}~\bibnamefont {Kn\"{o}bel}}, \bibinfo {author} {\bibfnamefont
  {S.}~\bibnamefont {Krupko}}, \bibinfo {author} {\bibfnamefont
  {M.}~\bibnamefont {Kuich}}, \bibinfo {author} {\bibfnamefont {Y.~A.}\
  \bibnamefont {Litvinov}}, \bibinfo {author} {\bibfnamefont {G.}~\bibnamefont
  {Marquinez-Dur\'{a}n}}, \bibinfo {author} {\bibfnamefont {I.}~\bibnamefont
  {Martel}}, \bibinfo {author} {\bibfnamefont {C.}~\bibnamefont {Mazzocchi}},
  \bibinfo {author} {\bibfnamefont {C.}~\bibnamefont {Nociforo}}, \bibinfo
  {author} {\bibfnamefont {A.~K.}\ \bibnamefont {Ord\'{u}z}}, \bibinfo {author}
  {\bibfnamefont {M.}~\bibnamefont {Pf\"{u}tzner}}, \bibinfo {author}
  {\bibfnamefont {S.}~\bibnamefont {Pietri}}, \bibinfo {author} {\bibfnamefont
  {M.}~\bibnamefont {Pomorski}}, \bibinfo {author} {\bibfnamefont
  {A.}~\bibnamefont {Prochazka}}, \bibinfo {author} {\bibfnamefont
  {S.}~\bibnamefont {Rymzhanova}}, \bibinfo {author} {\bibfnamefont {A.~M.}\
  \bibnamefont {S\'{a}nchez-Ben\'{i}tez}}, \bibinfo {author} {\bibfnamefont
  {C.}~\bibnamefont {Scheidenberger}}, \bibinfo {author} {\bibfnamefont
  {P.}~\bibnamefont {Sharov}}, \bibinfo {author} {\bibfnamefont
  {H.}~\bibnamefont {Simon}}, \bibinfo {author} {\bibfnamefont
  {B.}~\bibnamefont {Sitar}}, \bibinfo {author} {\bibfnamefont
  {R.}~\bibnamefont {Slepnev}}, \bibinfo {author} {\bibfnamefont
  {M.}~\bibnamefont {Stanoiu}}, \bibinfo {author} {\bibfnamefont
  {P.}~\bibnamefont {Strmen}}, \bibinfo {author} {\bibfnamefont
  {I.}~\bibnamefont {Szarka}}, \bibinfo {author} {\bibfnamefont
  {M.}~\bibnamefont {Takechi}}, \bibinfo {author} {\bibfnamefont {Y.~K.}\
  \bibnamefont {Tanaka}}, \bibinfo {author} {\bibfnamefont {H.}~\bibnamefont
  {Weick}}, \bibinfo {author} {\bibfnamefont {M.}~\bibnamefont {Winkler}},
  \bibinfo {author} {\bibfnamefont {J.~S.}\ \bibnamefont {Winfield}}, \ and\
  \bibinfo {author} {\bibfnamefont {M.~V.}\ \bibnamefont {Zhukov}},\
  }\href@noop {} {\bibfield  {journal} {\bibinfo  {journal} {Phys. Rev. Lett.}\
  }\textbf {\bibinfo {volume} {115}},\ \bibinfo {pages} {202501} (\bibinfo
  {year} {2015})}\BibitemShut {NoStop}%
\bibitem [{\citenamefont {Golubkova}\ \emph {et~al.}(2016)\citenamefont
  {Golubkova}, \citenamefont {Xu}, \citenamefont {Grigorenko}, \citenamefont
  {Mukha}, \citenamefont {Scheidenberger},\ and\ \citenamefont
  {Zhukov}}]{Golubkova:2016}%
  \BibitemOpen
  \bibfield  {author} {\bibinfo {author} {\bibfnamefont {T.}~\bibnamefont
  {Golubkova}}, \bibinfo {author} {\bibfnamefont {X.-D.}\ \bibnamefont {Xu}},
  \bibinfo {author} {\bibfnamefont {L.}~\bibnamefont {Grigorenko}}, \bibinfo
  {author} {\bibfnamefont {I.}~\bibnamefont {Mukha}}, \bibinfo {author}
  {\bibfnamefont {C.}~\bibnamefont {Scheidenberger}}, \ and\ \bibinfo {author}
  {\bibfnamefont {M.}~\bibnamefont {Zhukov}},\ }\href {\doibase
  http://dx.doi.org/10.1016/j.physletb.2016.09.034} {\bibfield  {journal}
  {\bibinfo  {journal} {Physics Letters B}\ }\textbf {\bibinfo {volume}
  {762}},\ \bibinfo {pages} {263 } (\bibinfo {year} {2016})}\BibitemShut
  {NoStop}%
\bibitem [{\citenamefont {Xu}\ \emph {et~al.}(2018)\citenamefont {Xu},
  \citenamefont {Mukha}, \citenamefont {Grigorenko}, \citenamefont
  {Scheidenberger}, \citenamefont {Acosta}, \citenamefont {Casarejos},
  \citenamefont {Chudoba}, \citenamefont {Ciemny}, \citenamefont {Dominik},
  \citenamefont {Du\'enas-D\'{\i}az}, \citenamefont {Dunin}, \citenamefont
  {Espino}, \citenamefont {Estrad\'e}, \citenamefont {Farinon}, \citenamefont
  {Fomichev}, \citenamefont {Geissel}, \citenamefont {Golubkova}, \citenamefont
  {Gorshkov}, \citenamefont {Janas}, \citenamefont {Kami\ifmmode~\acute{n}\else
  \'{n}\fi{}ski}, \citenamefont {Kiselev}, \citenamefont {Kn\"obel},
  \citenamefont {Krupko}, \citenamefont {Kuich}, \citenamefont {Litvinov},
  \citenamefont {Marquinez-Dur\'an}, \citenamefont {Martel}, \citenamefont
  {Mazzocchi}, \citenamefont {Nociforo}, \citenamefont {Ord\'uz}, \citenamefont
  {Pf\"utzner}, \citenamefont {Pietri}, \citenamefont {Pomorski}, \citenamefont
  {Prochazka}, \citenamefont {Rymzhanova}, \citenamefont
  {S\'anchez-Ben\'{\i}tez}, \citenamefont {Sharov}, \citenamefont {Simon},
  \citenamefont {Sitar}, \citenamefont {Slepnev}, \citenamefont {Stanoiu},
  \citenamefont {Strmen}, \citenamefont {Szarka}, \citenamefont {Takechi},
  \citenamefont {Tanaka}, \citenamefont {Weick}, \citenamefont {Winkler},\ and\
  \citenamefont {Winfield}}]{Xu:2018}%
  \BibitemOpen
  \bibfield  {author} {\bibinfo {author} {\bibfnamefont {X.-D.}\ \bibnamefont
  {Xu}}, \bibinfo {author} {\bibfnamefont {I.}~\bibnamefont {Mukha}}, \bibinfo
  {author} {\bibfnamefont {L.~V.}\ \bibnamefont {Grigorenko}}, \bibinfo
  {author} {\bibfnamefont {C.}~\bibnamefont {Scheidenberger}}, \bibinfo
  {author} {\bibfnamefont {L.}~\bibnamefont {Acosta}}, \bibinfo {author}
  {\bibfnamefont {E.}~\bibnamefont {Casarejos}}, \bibinfo {author}
  {\bibfnamefont {V.}~\bibnamefont {Chudoba}}, \bibinfo {author} {\bibfnamefont
  {A.~A.}\ \bibnamefont {Ciemny}}, \bibinfo {author} {\bibfnamefont
  {W.}~\bibnamefont {Dominik}}, \bibinfo {author} {\bibfnamefont
  {J.}~\bibnamefont {Du\'enas-D\'{\i}az}}, \bibinfo {author} {\bibfnamefont
  {V.}~\bibnamefont {Dunin}}, \bibinfo {author} {\bibfnamefont {J.~M.}\
  \bibnamefont {Espino}}, \bibinfo {author} {\bibfnamefont {A.}~\bibnamefont
  {Estrad\'e}}, \bibinfo {author} {\bibfnamefont {F.}~\bibnamefont {Farinon}},
  \bibinfo {author} {\bibfnamefont {A.}~\bibnamefont {Fomichev}}, \bibinfo
  {author} {\bibfnamefont {H.}~\bibnamefont {Geissel}}, \bibinfo {author}
  {\bibfnamefont {T.~A.}\ \bibnamefont {Golubkova}}, \bibinfo {author}
  {\bibfnamefont {A.}~\bibnamefont {Gorshkov}}, \bibinfo {author}
  {\bibfnamefont {Z.}~\bibnamefont {Janas}}, \bibinfo {author} {\bibfnamefont
  {G.}~\bibnamefont {Kami\ifmmode~\acute{n}\else \'{n}\fi{}ski}}, \bibinfo
  {author} {\bibfnamefont {O.}~\bibnamefont {Kiselev}}, \bibinfo {author}
  {\bibfnamefont {R.}~\bibnamefont {Kn\"obel}}, \bibinfo {author}
  {\bibfnamefont {S.}~\bibnamefont {Krupko}}, \bibinfo {author} {\bibfnamefont
  {M.}~\bibnamefont {Kuich}}, \bibinfo {author} {\bibfnamefont {Y.~A.}\
  \bibnamefont {Litvinov}}, \bibinfo {author} {\bibfnamefont {G.}~\bibnamefont
  {Marquinez-Dur\'an}}, \bibinfo {author} {\bibfnamefont {I.}~\bibnamefont
  {Martel}}, \bibinfo {author} {\bibfnamefont {C.}~\bibnamefont {Mazzocchi}},
  \bibinfo {author} {\bibfnamefont {C.}~\bibnamefont {Nociforo}}, \bibinfo
  {author} {\bibfnamefont {A.~K.}\ \bibnamefont {Ord\'uz}}, \bibinfo {author}
  {\bibfnamefont {M.}~\bibnamefont {Pf\"utzner}}, \bibinfo {author}
  {\bibfnamefont {S.}~\bibnamefont {Pietri}}, \bibinfo {author} {\bibfnamefont
  {M.}~\bibnamefont {Pomorski}}, \bibinfo {author} {\bibfnamefont
  {A.}~\bibnamefont {Prochazka}}, \bibinfo {author} {\bibfnamefont
  {S.}~\bibnamefont {Rymzhanova}}, \bibinfo {author} {\bibfnamefont {A.~M.}\
  \bibnamefont {S\'anchez-Ben\'{\i}tez}}, \bibinfo {author} {\bibfnamefont
  {P.}~\bibnamefont {Sharov}}, \bibinfo {author} {\bibfnamefont
  {H.}~\bibnamefont {Simon}}, \bibinfo {author} {\bibfnamefont
  {B.}~\bibnamefont {Sitar}}, \bibinfo {author} {\bibfnamefont
  {R.}~\bibnamefont {Slepnev}}, \bibinfo {author} {\bibfnamefont
  {M.}~\bibnamefont {Stanoiu}}, \bibinfo {author} {\bibfnamefont
  {P.}~\bibnamefont {Strmen}}, \bibinfo {author} {\bibfnamefont
  {I.}~\bibnamefont {Szarka}}, \bibinfo {author} {\bibfnamefont
  {M.}~\bibnamefont {Takechi}}, \bibinfo {author} {\bibfnamefont {Y.~K.}\
  \bibnamefont {Tanaka}}, \bibinfo {author} {\bibfnamefont {H.}~\bibnamefont
  {Weick}}, \bibinfo {author} {\bibfnamefont {M.}~\bibnamefont {Winkler}}, \
  and\ \bibinfo {author} {\bibfnamefont {J.~S.}\ \bibnamefont {Winfield}},\
  }\href {\doibase 10.1103/PhysRevC.97.034305} {\bibfield  {journal} {\bibinfo
  {journal} {Phys. Rev. C}\ }\textbf {\bibinfo {volume} {97}},\ \bibinfo
  {pages} {034305} (\bibinfo {year} {2018})}\BibitemShut {NoStop}%
\bibitem [{\citenamefont {{National Nuclear Data Center}}()}]{www:nndc2}%
  \BibitemOpen
  \bibfield  {author} {\bibinfo {author} {\bibnamefont {{National Nuclear Data
  Center}}},\ }\href@noop {} {}\bibinfo {howpublished}
  {\url{http://www.nndc.bnl.gov/}}\BibitemShut {NoStop}%
\bibitem [{\citenamefont {Audi}\ \emph {et~al.}(2014)\citenamefont {Audi},
  \citenamefont {Wang}, \citenamefont {Wapstra}, \citenamefont {MacCormick},\
  and\ \citenamefont {Xu}}]{Audi:2014}%
  \BibitemOpen
  \bibfield  {author} {\bibinfo {author} {\bibfnamefont {G.}~\bibnamefont
  {Audi}}, \bibinfo {author} {\bibfnamefont {M.}~\bibnamefont {Wang}}, \bibinfo
  {author} {\bibfnamefont {A.}~\bibnamefont {Wapstra}}, \bibinfo {author}
  {\bibfnamefont {F.~K.~M.}\ \bibnamefont {MacCormick}}, \ and\ \bibinfo
  {author} {\bibfnamefont {X.}~\bibnamefont {Xu}},\ }\href@noop {} {\bibfield
  {journal} {\bibinfo  {journal} {Nucl. Data Sheets}\ }\textbf {\bibinfo
  {volume} {120}},\ \bibinfo {pages} {1} (\bibinfo {year} {2014})},\ \bibinfo
  {note} {compilation A=1-270; atomic masses, other properties.}\BibitemShut
  {Stop}%
\bibitem [{\citenamefont {Tian}\ \emph {et~al.}(2013)\citenamefont {Tian},
  \citenamefont {Wang}, \citenamefont {Li},\ and\ \citenamefont
  {Li}}]{Tian:2013}%
  \BibitemOpen
  \bibfield  {author} {\bibinfo {author} {\bibfnamefont {J.}~\bibnamefont
  {Tian}}, \bibinfo {author} {\bibfnamefont {N.}~\bibnamefont {Wang}}, \bibinfo
  {author} {\bibfnamefont {C.}~\bibnamefont {Li}}, \ and\ \bibinfo {author}
  {\bibfnamefont {J.}~\bibnamefont {Li}},\ }\href {\doibase
  10.1103/PhysRevC.87.014313} {\bibfield  {journal} {\bibinfo  {journal} {Phys.
  Rev. C}\ }\textbf {\bibinfo {volume} {87}},\ \bibinfo {pages} {014313}
  (\bibinfo {year} {2013})}\BibitemShut {NoStop}%
\bibitem [{\citenamefont {Pf\"utzner}\ \emph {et~al.}(2012)\citenamefont
  {Pf\"utzner}, \citenamefont {Karny}, \citenamefont {Grigorenko},\ and\
  \citenamefont {Riisager}}]{Pfutzner:2012}%
  \BibitemOpen
  \bibfield  {author} {\bibinfo {author} {\bibfnamefont {M.}~\bibnamefont
  {Pf\"utzner}}, \bibinfo {author} {\bibfnamefont {M.}~\bibnamefont {Karny}},
  \bibinfo {author} {\bibfnamefont {L.~V.}\ \bibnamefont {Grigorenko}}, \ and\
  \bibinfo {author} {\bibfnamefont {K.}~\bibnamefont {Riisager}},\ }\href
  {\doibase 10.1103/RevModPhys.84.567} {\bibfield  {journal} {\bibinfo
  {journal} {Rev. Mod. Phys.}\ }\textbf {\bibinfo {volume} {84}},\ \bibinfo
  {pages} {567} (\bibinfo {year} {2012})}\BibitemShut {NoStop}%
\bibitem [{\citenamefont {Sharov}\ \emph {et~al.}(2014)\citenamefont {Sharov},
  \citenamefont {Egorova},\ and\ \citenamefont {Grigorenko}}]{Sharov:2014}%
  \BibitemOpen
  \bibfield  {author} {\bibinfo {author} {\bibfnamefont {P.~G.}\ \bibnamefont
  {Sharov}}, \bibinfo {author} {\bibfnamefont {I.~A.}\ \bibnamefont {Egorova}},
  \ and\ \bibinfo {author} {\bibfnamefont {L.~V.}\ \bibnamefont {Grigorenko}},\
  }\href {\doibase 10.1103/PhysRevC.90.024610} {\bibfield  {journal} {\bibinfo
  {journal} {Phys. Rev. C}\ }\textbf {\bibinfo {volume} {90}},\ \bibinfo
  {pages} {024610} (\bibinfo {year} {2014})}\BibitemShut {NoStop}%
\bibitem [{\citenamefont {Grigorenko}\ \emph {et~al.}(2004)\citenamefont
  {Grigorenko}, \citenamefont {Timofeyuk},\ and\ \citenamefont
  {Zhukov}}]{Grigorenko:2004}%
  \BibitemOpen
  \bibfield  {author} {\bibinfo {author} {\bibfnamefont {L.~V.}\ \bibnamefont
  {Grigorenko}}, \bibinfo {author} {\bibfnamefont {N.~K.}\ \bibnamefont
  {Timofeyuk}}, \ and\ \bibinfo {author} {\bibfnamefont {M.~V.}\ \bibnamefont
  {Zhukov}},\ }\href@noop {} {\bibfield  {journal} {\bibinfo  {journal} {Eur.
  Phys. J. A}\ }\textbf {\bibinfo {volume} {19}},\ \bibinfo {pages} {187}
  (\bibinfo {year} {2004})}\BibitemShut {NoStop}%
\bibitem [{\citenamefont {{Technical Design Report of the EXPERT setup for the
  Super-FRS Experiment Collaboration}}()}]{www:expert-tdr}%
  \BibitemOpen
  \bibfield  {author} {\bibinfo {author} {\bibnamefont {{Technical Design
  Report of the EXPERT setup for the Super-FRS Experiment Collaboration}}},\
  }\href@noop {} {}\bibinfo {howpublished}
  {\url{http://edms.cern.ch/document/1865700}}\BibitemShut {NoStop}%
\bibitem [{\citenamefont {Aysto}\ \emph {et~al.}(2016)\citenamefont {Aysto},
  \citenamefont {Behr}, \citenamefont {Benlliure}, \citenamefont {Bracco},
  \citenamefont {Egelhof}, \citenamefont {Fomichev}, \citenamefont {Gales},
  \citenamefont {Geissel}, \citenamefont {Grahn}, \citenamefont {Grigorenko},
  \citenamefont {Harakeh}, \citenamefont {Hayano}, \citenamefont {Heinz},
  \citenamefont {Itahashi}, \citenamefont {Jokinen}, \citenamefont
  {Kalantar-Nayestanaki}, \citenamefont {Kanungo}, \citenamefont {Lenske},
  \citenamefont {Mukha}, \citenamefont {Munzenberg}, \citenamefont {Nociforo},
  \citenamefont {Ong}, \citenamefont {Pietri}, \citenamefont {Pfutzner},
  \citenamefont {Plass}, \citenamefont {Prochazka}, \citenamefont
  {Purushothaman}, \citenamefont {Saito}, \citenamefont {Scheidenberger},
  \citenamefont {Simon}, \citenamefont {Tanihata}, \citenamefont {Terashima},
  \citenamefont {Toki}, \citenamefont {Trache}, \citenamefont {Weick},
  \citenamefont {Winfield}, \citenamefont {Winkler},\ and\ \citenamefont
  {Zamfir}}]{Aysto:2016}%
  \BibitemOpen
  \bibfield  {author} {\bibinfo {author} {\bibfnamefont {J.}~\bibnamefont
  {Aysto}}, \bibinfo {author} {\bibfnamefont {K.-H.}\ \bibnamefont {Behr}},
  \bibinfo {author} {\bibfnamefont {J.}~\bibnamefont {Benlliure}}, \bibinfo
  {author} {\bibfnamefont {A.}~\bibnamefont {Bracco}}, \bibinfo {author}
  {\bibfnamefont {P.}~\bibnamefont {Egelhof}}, \bibinfo {author} {\bibfnamefont
  {A.}~\bibnamefont {Fomichev}}, \bibinfo {author} {\bibfnamefont
  {S.}~\bibnamefont {Gales}}, \bibinfo {author} {\bibfnamefont
  {H.}~\bibnamefont {Geissel}}, \bibinfo {author} {\bibfnamefont
  {T.}~\bibnamefont {Grahn}}, \bibinfo {author} {\bibfnamefont
  {L.}~\bibnamefont {Grigorenko}}, \bibinfo {author} {\bibfnamefont
  {M.}~\bibnamefont {Harakeh}}, \bibinfo {author} {\bibfnamefont
  {R.}~\bibnamefont {Hayano}}, \bibinfo {author} {\bibfnamefont
  {S.}~\bibnamefont {Heinz}}, \bibinfo {author} {\bibfnamefont
  {K.}~\bibnamefont {Itahashi}}, \bibinfo {author} {\bibfnamefont
  {A.}~\bibnamefont {Jokinen}}, \bibinfo {author} {\bibfnamefont
  {N.}~\bibnamefont {Kalantar-Nayestanaki}}, \bibinfo {author} {\bibfnamefont
  {R.}~\bibnamefont {Kanungo}}, \bibinfo {author} {\bibfnamefont
  {H.}~\bibnamefont {Lenske}}, \bibinfo {author} {\bibfnamefont
  {I.}~\bibnamefont {Mukha}}, \bibinfo {author} {\bibfnamefont
  {G.}~\bibnamefont {Munzenberg}}, \bibinfo {author} {\bibfnamefont
  {C.}~\bibnamefont {Nociforo}}, \bibinfo {author} {\bibfnamefont
  {H.}~\bibnamefont {Ong}}, \bibinfo {author} {\bibfnamefont {S.}~\bibnamefont
  {Pietri}}, \bibinfo {author} {\bibfnamefont {M.}~\bibnamefont {Pfutzner}},
  \bibinfo {author} {\bibfnamefont {W.}~\bibnamefont {Plass}}, \bibinfo
  {author} {\bibfnamefont {A.}~\bibnamefont {Prochazka}}, \bibinfo {author}
  {\bibfnamefont {S.}~\bibnamefont {Purushothaman}}, \bibinfo {author}
  {\bibfnamefont {T.}~\bibnamefont {Saito}}, \bibinfo {author} {\bibfnamefont
  {C.}~\bibnamefont {Scheidenberger}}, \bibinfo {author} {\bibfnamefont
  {H.}~\bibnamefont {Simon}}, \bibinfo {author} {\bibfnamefont
  {I.}~\bibnamefont {Tanihata}}, \bibinfo {author} {\bibfnamefont
  {S.}~\bibnamefont {Terashima}}, \bibinfo {author} {\bibfnamefont
  {H.}~\bibnamefont {Toki}}, \bibinfo {author} {\bibfnamefont {L.}~\bibnamefont
  {Trache}}, \bibinfo {author} {\bibfnamefont {H.}~\bibnamefont {Weick}},
  \bibinfo {author} {\bibfnamefont {J.}~\bibnamefont {Winfield}}, \bibinfo
  {author} {\bibfnamefont {M.}~\bibnamefont {Winkler}}, \ and\ \bibinfo
  {author} {\bibfnamefont {V.}~\bibnamefont {Zamfir}},\ }\href {\doibase
  https://doi.org/10.1016/j.nimb.2016.02.042} {\bibfield  {journal} {\bibinfo
  {journal} {Nucl. Instr. Meth. in Phys. Res., B}\ }\textbf {\bibinfo {volume}
  {376}},\ \bibinfo {pages} {111 } (\bibinfo {year} {2016})},\ \bibinfo {note}
  {proceedings of the XVIIth International Conference on Electromagnetic
  Isotope Separators and Related Topics (EMIS2015), Grand Rapids, MI, U.S.A.,
  11-15 May 2015}\BibitemShut {NoStop}%
\bibitem [{\citenamefont {Grigorenko}\ and\ \citenamefont
  {Zhukov}(2007{\natexlab{a}})}]{Grigorenko:2007}%
  \BibitemOpen
  \bibfield  {author} {\bibinfo {author} {\bibfnamefont {L.~V.}\ \bibnamefont
  {Grigorenko}}\ and\ \bibinfo {author} {\bibfnamefont {M.~V.}\ \bibnamefont
  {Zhukov}},\ }\href {\doibase 10.1103/PhysRevC.76.014008} {\bibfield
  {journal} {\bibinfo  {journal} {Phys. Rev. C}\ }\textbf {\bibinfo {volume}
  {76}},\ \bibinfo {pages} {014008} (\bibinfo {year}
  {2007}{\natexlab{a}})}\BibitemShut {NoStop}%
\bibitem [{\citenamefont {Grigorenko}\ and\ \citenamefont
  {Zhukov}(2007{\natexlab{b}})}]{Grigorenko:2007a}%
  \BibitemOpen
  \bibfield  {author} {\bibinfo {author} {\bibfnamefont {L.~V.}\ \bibnamefont
  {Grigorenko}}\ and\ \bibinfo {author} {\bibfnamefont {M.~V.}\ \bibnamefont
  {Zhukov}},\ }\href@noop {} {\bibfield  {journal} {\bibinfo  {journal} {Phys.
  Rev. C}\ }\textbf {\bibinfo {volume} {76}},\ \bibinfo {pages} {014009}
  (\bibinfo {year} {2007}{\natexlab{b}})}\BibitemShut {NoStop}%
\bibitem [{\citenamefont {Grigorenko}\ \emph {et~al.}(2013)\citenamefont
  {Grigorenko}, \citenamefont {Mukha},\ and\ \citenamefont
  {Zhukov}}]{Grigorenko:2013}%
  \BibitemOpen
  \bibfield  {author} {\bibinfo {author} {\bibfnamefont {L.~V.}\ \bibnamefont
  {Grigorenko}}, \bibinfo {author} {\bibfnamefont {I.~G.}\ \bibnamefont
  {Mukha}}, \ and\ \bibinfo {author} {\bibfnamefont {M.~V.}\ \bibnamefont
  {Zhukov}},\ }\href {\doibase 10.1103/PhysRevLett.111.042501} {\bibfield
  {journal} {\bibinfo  {journal} {Phys. Rev. Lett.}\ }\textbf {\bibinfo
  {volume} {111}},\ \bibinfo {pages} {042501} (\bibinfo {year}
  {2013})}\BibitemShut {NoStop}%
\bibitem [{\citenamefont {Grigorenko}\ and\ \citenamefont
  {Zhukov}(2015)}]{Grigorenko:2015b}%
  \BibitemOpen
  \bibfield  {author} {\bibinfo {author} {\bibfnamefont {L.~V.}\ \bibnamefont
  {Grigorenko}}\ and\ \bibinfo {author} {\bibfnamefont {M.~V.}\ \bibnamefont
  {Zhukov}},\ }\href@noop {} {\bibfield  {journal} {\bibinfo  {journal} {Phys.
  Rev. C}\ }\textbf {\bibinfo {volume} {91}},\ \bibinfo {pages} {064617}
  (\bibinfo {year} {2015})},\ \bibinfo {note} {[ArXiv:1503.03186]}\BibitemShut
  {NoStop}%
\bibitem [{\citenamefont {Ehrman}(1951)}]{Ehrman:1951}%
  \BibitemOpen
  \bibfield  {author} {\bibinfo {author} {\bibfnamefont {J.~B.}\ \bibnamefont
  {Ehrman}},\ }\href {\doibase 10.1103/PhysRev.81.412} {\bibfield  {journal}
  {\bibinfo  {journal} {Phys. Rev.}\ }\textbf {\bibinfo {volume} {81}},\
  \bibinfo {pages} {412} (\bibinfo {year} {1951})}\BibitemShut {NoStop}%
\bibitem [{\citenamefont {Thomas}(1952)}]{Thomas:1952}%
  \BibitemOpen
  \bibfield  {author} {\bibinfo {author} {\bibfnamefont {R.~G.}\ \bibnamefont
  {Thomas}},\ }\href {\doibase 10.1103/PhysRev.88.1109} {\bibfield  {journal}
  {\bibinfo  {journal} {Phys. Rev.}\ }\textbf {\bibinfo {volume} {88}},\
  \bibinfo {pages} {1109} (\bibinfo {year} {1952})}\BibitemShut {NoStop}%
\bibitem [{\citenamefont {Hoffman}\ \emph {et~al.}(2008)\citenamefont
  {Hoffman}, \citenamefont {Baumann}, \citenamefont {Bazin}, \citenamefont
  {Brown}, \citenamefont {Christian}, \citenamefont {DeYoung}, \citenamefont
  {Finck}, \citenamefont {Frank}, \citenamefont {Hinnefeld}, \citenamefont
  {Howes}, \citenamefont {Mears}, \citenamefont {Mosby}, \citenamefont {Mosby},
  \citenamefont {Reith}, \citenamefont {Rizzo}, \citenamefont {Rogers},
  \citenamefont {Peaslee}, \citenamefont {Peters}, \citenamefont {Schiller},
  \citenamefont {Scott}, \citenamefont {Tabor}, \citenamefont {Thoennessen},
  \citenamefont {Voss},\ and\ \citenamefont {Williams}}]{Hoffman:2008}%
  \BibitemOpen
  \bibfield  {author} {\bibinfo {author} {\bibfnamefont {C.~R.}\ \bibnamefont
  {Hoffman}}, \bibinfo {author} {\bibfnamefont {T.}~\bibnamefont {Baumann}},
  \bibinfo {author} {\bibfnamefont {D.}~\bibnamefont {Bazin}}, \bibinfo
  {author} {\bibfnamefont {J.}~\bibnamefont {Brown}}, \bibinfo {author}
  {\bibfnamefont {G.}~\bibnamefont {Christian}}, \bibinfo {author}
  {\bibfnamefont {P.~A.}\ \bibnamefont {DeYoung}}, \bibinfo {author}
  {\bibfnamefont {J.~E.}\ \bibnamefont {Finck}}, \bibinfo {author}
  {\bibfnamefont {N.}~\bibnamefont {Frank}}, \bibinfo {author} {\bibfnamefont
  {J.}~\bibnamefont {Hinnefeld}}, \bibinfo {author} {\bibfnamefont
  {R.}~\bibnamefont {Howes}}, \bibinfo {author} {\bibfnamefont
  {P.}~\bibnamefont {Mears}}, \bibinfo {author} {\bibfnamefont
  {E.}~\bibnamefont {Mosby}}, \bibinfo {author} {\bibfnamefont
  {S.}~\bibnamefont {Mosby}}, \bibinfo {author} {\bibfnamefont
  {J.}~\bibnamefont {Reith}}, \bibinfo {author} {\bibfnamefont
  {B.}~\bibnamefont {Rizzo}}, \bibinfo {author} {\bibfnamefont {W.~F.}\
  \bibnamefont {Rogers}}, \bibinfo {author} {\bibfnamefont {G.}~\bibnamefont
  {Peaslee}}, \bibinfo {author} {\bibfnamefont {W.~A.}\ \bibnamefont {Peters}},
  \bibinfo {author} {\bibfnamefont {A.}~\bibnamefont {Schiller}}, \bibinfo
  {author} {\bibfnamefont {M.~J.}\ \bibnamefont {Scott}}, \bibinfo {author}
  {\bibfnamefont {S.~L.}\ \bibnamefont {Tabor}}, \bibinfo {author}
  {\bibfnamefont {M.}~\bibnamefont {Thoennessen}}, \bibinfo {author}
  {\bibfnamefont {P.~J.}\ \bibnamefont {Voss}}, \ and\ \bibinfo {author}
  {\bibfnamefont {T.}~\bibnamefont {Williams}},\ }\href {\doibase
  10.1103/PhysRevLett.100.152502} {\bibfield  {journal} {\bibinfo  {journal}
  {Phys. Rev. Lett.}\ }\textbf {\bibinfo {volume} {100}},\ \bibinfo {pages}
  {152502} (\bibinfo {year} {2008})}\BibitemShut {NoStop}%
\bibitem [{\citenamefont {Kondo}\ \emph {et~al.}(2016)\citenamefont {Kondo},
  \citenamefont {Nakamura}, \citenamefont {Tanaka}, \citenamefont {Minakata},
  \citenamefont {Ogoshi}, \citenamefont {Orr}, \citenamefont {Achouri},
  \citenamefont {Aumann}, \citenamefont {Baba}, \citenamefont {Delaunay},
  \citenamefont {Doornenbal}, \citenamefont {Fukuda}, \citenamefont {Gibelin},
  \citenamefont {Hwang}, \citenamefont {Inabe}, \citenamefont {Isobe},
  \citenamefont {Kameda}, \citenamefont {Kanno}, \citenamefont {Kim},
  \citenamefont {Kobayashi}, \citenamefont {Kobayashi}, \citenamefont {Kubo},
  \citenamefont {Leblond}, \citenamefont {Lee}, \citenamefont {Marqu\'es},
  \citenamefont {Motobayashi}, \citenamefont {Murai}, \citenamefont {Murakami},
  \citenamefont {Muto}, \citenamefont {Nakashima}, \citenamefont {Nakatsuka},
  \citenamefont {Navin}, \citenamefont {Nishi}, \citenamefont {Otsu},
  \citenamefont {Sato}, \citenamefont {Satou}, \citenamefont {Shimizu},
  \citenamefont {Suzuki}, \citenamefont {Takahashi}, \citenamefont {Takeda},
  \citenamefont {Takeuchi}, \citenamefont {Togano}, \citenamefont {Tuff},
  \citenamefont {Vandebrouck},\ and\ \citenamefont {Yoneda}}]{Kondo:2016}%
  \BibitemOpen
  \bibfield  {author} {\bibinfo {author} {\bibfnamefont {Y.}~\bibnamefont
  {Kondo}}, \bibinfo {author} {\bibfnamefont {T.}~\bibnamefont {Nakamura}},
  \bibinfo {author} {\bibfnamefont {R.}~\bibnamefont {Tanaka}}, \bibinfo
  {author} {\bibfnamefont {R.}~\bibnamefont {Minakata}}, \bibinfo {author}
  {\bibfnamefont {S.}~\bibnamefont {Ogoshi}}, \bibinfo {author} {\bibfnamefont
  {N.~A.}\ \bibnamefont {Orr}}, \bibinfo {author} {\bibfnamefont {N.~L.}\
  \bibnamefont {Achouri}}, \bibinfo {author} {\bibfnamefont {T.}~\bibnamefont
  {Aumann}}, \bibinfo {author} {\bibfnamefont {H.}~\bibnamefont {Baba}},
  \bibinfo {author} {\bibfnamefont {F.}~\bibnamefont {Delaunay}}, \bibinfo
  {author} {\bibfnamefont {P.}~\bibnamefont {Doornenbal}}, \bibinfo {author}
  {\bibfnamefont {N.}~\bibnamefont {Fukuda}}, \bibinfo {author} {\bibfnamefont
  {J.}~\bibnamefont {Gibelin}}, \bibinfo {author} {\bibfnamefont {J.~W.}\
  \bibnamefont {Hwang}}, \bibinfo {author} {\bibfnamefont {N.}~\bibnamefont
  {Inabe}}, \bibinfo {author} {\bibfnamefont {T.}~\bibnamefont {Isobe}},
  \bibinfo {author} {\bibfnamefont {D.}~\bibnamefont {Kameda}}, \bibinfo
  {author} {\bibfnamefont {D.}~\bibnamefont {Kanno}}, \bibinfo {author}
  {\bibfnamefont {S.}~\bibnamefont {Kim}}, \bibinfo {author} {\bibfnamefont
  {N.}~\bibnamefont {Kobayashi}}, \bibinfo {author} {\bibfnamefont
  {T.}~\bibnamefont {Kobayashi}}, \bibinfo {author} {\bibfnamefont
  {T.}~\bibnamefont {Kubo}}, \bibinfo {author} {\bibfnamefont {S.}~\bibnamefont
  {Leblond}}, \bibinfo {author} {\bibfnamefont {J.}~\bibnamefont {Lee}},
  \bibinfo {author} {\bibfnamefont {F.~M.}\ \bibnamefont {Marqu\'es}}, \bibinfo
  {author} {\bibfnamefont {T.}~\bibnamefont {Motobayashi}}, \bibinfo {author}
  {\bibfnamefont {D.}~\bibnamefont {Murai}}, \bibinfo {author} {\bibfnamefont
  {T.}~\bibnamefont {Murakami}}, \bibinfo {author} {\bibfnamefont
  {K.}~\bibnamefont {Muto}}, \bibinfo {author} {\bibfnamefont {T.}~\bibnamefont
  {Nakashima}}, \bibinfo {author} {\bibfnamefont {N.}~\bibnamefont
  {Nakatsuka}}, \bibinfo {author} {\bibfnamefont {A.}~\bibnamefont {Navin}},
  \bibinfo {author} {\bibfnamefont {S.}~\bibnamefont {Nishi}}, \bibinfo
  {author} {\bibfnamefont {H.}~\bibnamefont {Otsu}}, \bibinfo {author}
  {\bibfnamefont {H.}~\bibnamefont {Sato}}, \bibinfo {author} {\bibfnamefont
  {Y.}~\bibnamefont {Satou}}, \bibinfo {author} {\bibfnamefont
  {Y.}~\bibnamefont {Shimizu}}, \bibinfo {author} {\bibfnamefont
  {H.}~\bibnamefont {Suzuki}}, \bibinfo {author} {\bibfnamefont
  {K.}~\bibnamefont {Takahashi}}, \bibinfo {author} {\bibfnamefont
  {H.}~\bibnamefont {Takeda}}, \bibinfo {author} {\bibfnamefont
  {S.}~\bibnamefont {Takeuchi}}, \bibinfo {author} {\bibfnamefont
  {Y.}~\bibnamefont {Togano}}, \bibinfo {author} {\bibfnamefont {A.~G.}\
  \bibnamefont {Tuff}}, \bibinfo {author} {\bibfnamefont {M.}~\bibnamefont
  {Vandebrouck}}, \ and\ \bibinfo {author} {\bibfnamefont {K.}~\bibnamefont
  {Yoneda}},\ }\href {\doibase 10.1103/PhysRevLett.116.102503} {\bibfield
  {journal} {\bibinfo  {journal} {Phys. Rev. Lett.}\ }\textbf {\bibinfo
  {volume} {116}},\ \bibinfo {pages} {102503} (\bibinfo {year}
  {2016})}\BibitemShut {NoStop}%
\bibitem [{\citenamefont {Jones}\ \emph {et~al.}(2017)\citenamefont {Jones},
  \citenamefont {Fossez}, \citenamefont {Baumann}, \citenamefont {DeYoung},
  \citenamefont {Finck}, \citenamefont {Frank}, \citenamefont {Kuchera},
  \citenamefont {Michel}, \citenamefont {Nazarewicz}, \citenamefont {Rotureau},
  \citenamefont {Smith}, \citenamefont {Stephenson}, \citenamefont {Stiefel},
  \citenamefont {Thoennessen},\ and\ \citenamefont {Zegers}}]{Jones:2017}%
  \BibitemOpen
  \bibfield  {author} {\bibinfo {author} {\bibfnamefont {M.~D.}\ \bibnamefont
  {Jones}}, \bibinfo {author} {\bibfnamefont {K.}~\bibnamefont {Fossez}},
  \bibinfo {author} {\bibfnamefont {T.}~\bibnamefont {Baumann}}, \bibinfo
  {author} {\bibfnamefont {P.~A.}\ \bibnamefont {DeYoung}}, \bibinfo {author}
  {\bibfnamefont {J.~E.}\ \bibnamefont {Finck}}, \bibinfo {author}
  {\bibfnamefont {N.}~\bibnamefont {Frank}}, \bibinfo {author} {\bibfnamefont
  {A.~N.}\ \bibnamefont {Kuchera}}, \bibinfo {author} {\bibfnamefont
  {N.}~\bibnamefont {Michel}}, \bibinfo {author} {\bibfnamefont
  {W.}~\bibnamefont {Nazarewicz}}, \bibinfo {author} {\bibfnamefont
  {J.}~\bibnamefont {Rotureau}}, \bibinfo {author} {\bibfnamefont {J.~K.}\
  \bibnamefont {Smith}}, \bibinfo {author} {\bibfnamefont {S.~L.}\ \bibnamefont
  {Stephenson}}, \bibinfo {author} {\bibfnamefont {K.}~\bibnamefont {Stiefel}},
  \bibinfo {author} {\bibfnamefont {M.}~\bibnamefont {Thoennessen}}, \ and\
  \bibinfo {author} {\bibfnamefont {R.~G.~T.}\ \bibnamefont {Zegers}},\ }\href
  {\doibase 10.1103/PhysRevC.96.054322} {\bibfield  {journal} {\bibinfo
  {journal} {Phys. Rev. C}\ }\textbf {\bibinfo {volume} {96}},\ \bibinfo
  {pages} {054322} (\bibinfo {year} {2017})}\BibitemShut {NoStop}%
\bibitem [{\citenamefont {Cole}(1996)}]{Cole:1996}%
  \BibitemOpen
  \bibfield  {author} {\bibinfo {author} {\bibfnamefont {B.~J.}\ \bibnamefont
  {Cole}},\ }\href@noop {} {\bibfield  {journal} {\bibinfo  {journal} {Phys.
  Rev. C}\ }\textbf {\bibinfo {volume} {54}},\ \bibinfo {pages} {1240}
  (\bibinfo {year} {1996})}\BibitemShut {NoStop}%
\bibitem [{\citenamefont {Lis}\ \emph {et~al.}(2015)\citenamefont {Lis},
  \citenamefont {Mazzocchi}, \citenamefont {Dominik}, \citenamefont {Janas},
  \citenamefont {Pf\"utzner}, \citenamefont {Pomorski}, \citenamefont {Acosta},
  \citenamefont {Baraeva}, \citenamefont {Casarejos}, \citenamefont
  {Du\'enas-D\'{\i}az}, \citenamefont {Dunin}, \citenamefont {Espino},
  \citenamefont {Estrade}, \citenamefont {Farinon}, \citenamefont {Fomichev},
  \citenamefont {Geissel}, \citenamefont {Gorshkov}, \citenamefont
  {Kami\ifmmode~\acute{n}\else \'{n}\fi{}ski}, \citenamefont {Kiselev},
  \citenamefont {Kn\"obel}, \citenamefont {Krupko}, \citenamefont {Kuich},
  \citenamefont {Litvinov}, \citenamefont {Marquinez-Dur\'an}, \citenamefont
  {Martel}, \citenamefont {Mukha}, \citenamefont {Nociforo}, \citenamefont
  {Ord\'uz}, \citenamefont {Pietri}, \citenamefont {Prochazka}, \citenamefont
  {S\'anchez-Ben\'{\i}tez}, \citenamefont {Simon}, \citenamefont {Sitar},
  \citenamefont {Slepnev}, \citenamefont {Stanoiu}, \citenamefont {Strmen},
  \citenamefont {Szarka}, \citenamefont {Takechi}, \citenamefont {Tanaka},
  \citenamefont {Weick},\ and\ \citenamefont {Winfield}}]{Lis:2015}%
  \BibitemOpen
  \bibfield  {author} {\bibinfo {author} {\bibfnamefont {A.~A.}\ \bibnamefont
  {Lis}}, \bibinfo {author} {\bibfnamefont {C.}~\bibnamefont {Mazzocchi}},
  \bibinfo {author} {\bibfnamefont {W.}~\bibnamefont {Dominik}}, \bibinfo
  {author} {\bibfnamefont {Z.}~\bibnamefont {Janas}}, \bibinfo {author}
  {\bibfnamefont {M.}~\bibnamefont {Pf\"utzner}}, \bibinfo {author}
  {\bibfnamefont {M.}~\bibnamefont {Pomorski}}, \bibinfo {author}
  {\bibfnamefont {L.}~\bibnamefont {Acosta}}, \bibinfo {author} {\bibfnamefont
  {S.}~\bibnamefont {Baraeva}}, \bibinfo {author} {\bibfnamefont
  {E.}~\bibnamefont {Casarejos}}, \bibinfo {author} {\bibfnamefont
  {J.}~\bibnamefont {Du\'enas-D\'{\i}az}}, \bibinfo {author} {\bibfnamefont
  {V.}~\bibnamefont {Dunin}}, \bibinfo {author} {\bibfnamefont {J.~M.}\
  \bibnamefont {Espino}}, \bibinfo {author} {\bibfnamefont {A.}~\bibnamefont
  {Estrade}}, \bibinfo {author} {\bibfnamefont {F.}~\bibnamefont {Farinon}},
  \bibinfo {author} {\bibfnamefont {A.}~\bibnamefont {Fomichev}}, \bibinfo
  {author} {\bibfnamefont {H.}~\bibnamefont {Geissel}}, \bibinfo {author}
  {\bibfnamefont {A.}~\bibnamefont {Gorshkov}}, \bibinfo {author}
  {\bibfnamefont {G.}~\bibnamefont {Kami\ifmmode~\acute{n}\else
  \'{n}\fi{}ski}}, \bibinfo {author} {\bibfnamefont {O.}~\bibnamefont
  {Kiselev}}, \bibinfo {author} {\bibfnamefont {R.}~\bibnamefont {Kn\"obel}},
  \bibinfo {author} {\bibfnamefont {S.}~\bibnamefont {Krupko}}, \bibinfo
  {author} {\bibfnamefont {M.}~\bibnamefont {Kuich}}, \bibinfo {author}
  {\bibfnamefont {Y.~A.}\ \bibnamefont {Litvinov}}, \bibinfo {author}
  {\bibfnamefont {G.}~\bibnamefont {Marquinez-Dur\'an}}, \bibinfo {author}
  {\bibfnamefont {I.}~\bibnamefont {Martel}}, \bibinfo {author} {\bibfnamefont
  {I.}~\bibnamefont {Mukha}}, \bibinfo {author} {\bibfnamefont
  {C.}~\bibnamefont {Nociforo}}, \bibinfo {author} {\bibfnamefont {A.~K.}\
  \bibnamefont {Ord\'uz}}, \bibinfo {author} {\bibfnamefont {S.}~\bibnamefont
  {Pietri}}, \bibinfo {author} {\bibfnamefont {A.}~\bibnamefont {Prochazka}},
  \bibinfo {author} {\bibfnamefont {A.~M.}\ \bibnamefont
  {S\'anchez-Ben\'{\i}tez}}, \bibinfo {author} {\bibfnamefont {H.}~\bibnamefont
  {Simon}}, \bibinfo {author} {\bibfnamefont {B.}~\bibnamefont {Sitar}},
  \bibinfo {author} {\bibfnamefont {R.}~\bibnamefont {Slepnev}}, \bibinfo
  {author} {\bibfnamefont {M.}~\bibnamefont {Stanoiu}}, \bibinfo {author}
  {\bibfnamefont {P.}~\bibnamefont {Strmen}}, \bibinfo {author} {\bibfnamefont
  {I.}~\bibnamefont {Szarka}}, \bibinfo {author} {\bibfnamefont
  {M.}~\bibnamefont {Takechi}}, \bibinfo {author} {\bibfnamefont
  {Y.}~\bibnamefont {Tanaka}}, \bibinfo {author} {\bibfnamefont
  {H.}~\bibnamefont {Weick}}, \ and\ \bibinfo {author} {\bibfnamefont {J.~S.}\
  \bibnamefont {Winfield}},\ }\href {\doibase 10.1103/PhysRevC.91.064309}
  {\bibfield  {journal} {\bibinfo  {journal} {Phys. Rev. C}\ }\textbf {\bibinfo
  {volume} {91}},\ \bibinfo {pages} {064309} (\bibinfo {year}
  {2015})}\BibitemShut {NoStop}%
\end{thebibliography}%


\end{document}